\documentclass[aps,prd,groupedaddress,nofootinbib,twocolumn,longbibliography]{revtex4-2}
\usepackage{graphicx,amsmath,mathtools,amssymb}
\usepackage[svgnames]{xcolor}
\usepackage{tcolorbox}
\usepackage{amsthm}
\usepackage{graphicx}
\usepackage{color,soul}
\usepackage{soul}
\usepackage{bm}
\usepackage{subfigure}
\usepackage{booktabs}
\usepackage{bigdelim}
\usepackage{multirow}
\usepackage{hyperref} % For hyperlinks in the PDF
\usepackage{slashed}
\usepackage{orcidlink}
\hypersetup{
	colorlinks=true,
	citecolor=DarkBlue,  % <--- 试试这个，非常经典的深蓝
	linkcolor=DarkBlue,   % 内部链接（公式/图表）
	filecolor=DarkBlue,
	urlcolor=DarkBlue    % 网址也用深蓝
}

\allowdisplaybreaks

\begin{document}
	\title{Decoding $Z_c(4430)$ and $Z_c(4200)$: The role of $P$-wave charmed mesons}
	
	\author{Jian-Bo Cheng$^{1,2}$\,\orcidlink{0000-0003-1202-4344}}\email{jbcheng@pku.edu.cn}
	
	\author{Zi-Yang Lin$^{2}$\,\orcidlink{0000-0001-7887-9391}}\email{lzy\_15@pku.edu.cn}

	\author{Jun-Zhang Wang$^{3}$\,\orcidlink{0000-0002-3404-8569}}\email{wangjzh@cqu.edu.cn}

	\author{Shi-Lin Zhu$^{2}$\,\orcidlink{0000-0002-4055-6906}}\email{zhusl@pku.edu.cn}
	
	\affiliation{$^1$College of Science, China University of Petroleum, Qingdao, Shandong 266580, China\\
		$^2$School of Physics and Center of High Energy Physics, Peking University, Beijing 100871, China\\
		$^3$Department of Physics and Chongqing Key Laboratory for Strongly Coupled Physics, Chongqing University, Chongqing 401331, China
	}

%	\date{\today}

\begin{abstract}
	In this work, we perform a systematic investigation of the hidden-charm tetraquark states with $I^G(J^{PC})=1^+(1^{+-})$ within the hadronic molecular picture, placing particular emphasis on systems composed of an $S$-wave $(D, D^*)$ meson and a $P$-wave $(D_0^*(2300), D_1(2430), D_1(2420), D_2^*(2460))$ meson. Adopting the One-Boson Exchange potential, we solve the Schrödinger equation in momentum space via the Complex Scaling Method. A crucial feature of our approach is the rigorous treatment of the unstable nature of the $P$-wave constituents by incorporating three-body decay effects arising from self-energy corrections and the static limit approximation. Our results demonstrate that these three-body dynamics play a crucial role in determining the pole positions, specifically in reproducing the large decay widths observed experimentally. We identify several broad resonances in the $D^*\bar{D}_1(2420)$ and $D^*\bar{D}_2^*(2460)$ systems as candidates for the $Z_c(4430)$, while the significantly broader resonances in the $D\bar{D}_0^*(2300)$ and $D\bar{D}_1(2430)$ sectors are suggested as candidates for the $Z_c(4200)$. Focusing on the $D^*\bar{D}_2^*(2460)$ assignment as a specific case study, we further analyze the line shape of the $Z_c(4430)$ candidate using a Flatté-like parametrization with energy-dependent self-energy terms, providing predictions for its open-charm decay modes to guide future experimental searches.
\end{abstract}
	%\pacs{14.40.Rt, 12.39.Pn}
	\maketitle
	
	%%%%%%%%%%%%%%%%%%%%%%%%%%%%%%%%%%%%%%%%%%%%%%%%%%%%%%
	\section{Introduction}\label{sec:Introduction}

	The observation of the $X(3872)$ by the Belle Collaboration in 2003 \cite{choiObservationNarrowCharmoniumlike2003} marked the dawn of a new era in hadron physics, revealing a rich spectrum of exotic candidates that defy the conventional $c\bar{c}$ quark model. Following this, the discovery of the charged charmonium-like states, such as the $Z_c(3900)$ and $Z_c(4020)$ by the BESIII and Belle Collaborations 	\cite{BESIIICollaboration2013,BelleCollaboration2013,Ablikim2013}, provided unambiguous evidence for the existence of multiquark states with a minimal content of $c\bar{c}u\bar{d}$. More recently, the LHCb Collaboration reported the $T_{cc}^+$ \cite{Aaij2022,LHCbCollaboration2022e}, a double-charm tetraquark candidate lying just below the $D^{*+}D^0$ threshold, which further enriched the landscape of exotic hadrons. For a broader perspective, comprehensive reviews published in recent years offer further insights into this field \cite{Hosaka:2016pey,chenHiddencharmPentaquarkTetraquark2016,Chen:2016spr,Ali:2017jda,Esposito2017,Lebed:2016hpi,guoHadronicMolecules2018,liuPentaquarkTetraquarkStates2019,brambillaStatesExperimentalTheoretical2020,Lucha:2021mwx,dongSurveyHeavyHeavy2021,Brambilla:2021mpo,Meng:2022ozq,Chen:2022asf}.
	
	Given that the $Z_c(3900)$ and $Z_c(4020)$ lie remarkably close to the $D\bar{D}^*$ and $D^*\bar{D}^*$ thresholds, respectively, a leading theoretical interpretation is that they are candidates for $S$-wave hadronic molecules \cite{Molina2009d,Zhang2011,Sun2011,Ozpineci2013,He2013,Guo2013,Dong2013,Wang2014a,He2014,Aceti2014,Wang2014,Karliner2015,Baru2019,Wang2020,Ding2020,Dai2022}, although their exact nature continues to be a subject of active research. While the low-lying $S$-wave molecular sector is relatively well-established, the nature of the higher-mass excited states remains an intriguing topic. A very recent experimental study \cite{Collaboration2025} reported updated observations of the $Z_c(4430)$, determining its mass and width as:
	\begin{align}
		M &= 4.452\pm0.016^{+0.055}_{-0.033}\ \text{GeV}, \nonumber\\
		\Gamma &= 0.174\pm0.019^{+0.083}_{-0.020} \ \text{GeV}. \label{LHCb 1}
	\end{align}
	This study also analyzed the detailed structure and corresponding decay modes of the state.  Historically, the $Z_c(4430)$ has been a focal point of exotic hadron research for over a decade. It was first observed by the Belle Collaboration in the $\psi(2S)\pi^+$ invariant mass spectrum \cite{Choi2008a} and subsequently confirmed by LHCb with high significance \cite{Aaij2014}, which determined its preferred quantum numbers to be $J^P=1^+$. Unlike the $Z_c(3900)$, the $Z_c(4430)$ is characterized by a significantly broader width and a mass lying far above the ground-state open-charm thresholds (such as $D\bar D^*$), complicating its theoretical interpretation.

	The prevailing theoretical perspective interprets the $Z_c(4430)$ as a hadronic molecule composed of two charmed mesons. In early works \cite{Liu2008,Liu2008a}, the authors explored the possibility of $Z_c(4430)$ as an $S$-wave  $D^*\bar{D}_1(2430)$/$D^*\bar{D}_1(2420)$ molecular state. Utilizing the One-Boson Exchange (OBE) model, Refs. \cite{He2016,He2017} further investigated the $P$-wave $D^*\bar{D}_1(2420)$ molecular scenario. Ma et al. \cite{Ma2014} studied the decay processes considering both the $S$-wave $\bar{D} D^*(2600)$ and the $P$-wave excited $D^*\bar{D}_2^*(2460)$ molecular configurations. 
	Furthermore, assuming a $\bar{D} D^*(2600)$ structure, Liu et al. \cite{Liu2014} analyzed the decay mechanisms, specifically comparing the $\psi(2S)\pi$ and $J/\psi\pi$ channels. The production mechanisms of this state have also been investigated in Refs. \cite{Liu2008b,Guerrieri2014}.
	Beyond the molecular picture, alternative explanations have been proposed, including the compact tetraquark structure \cite{Maiani2008,Ebert2008,Bracco2009,Wang2010,Maiani2014,Deng2015,Wang2015,Agaev2017,Azizi2020}, triangle singularities \cite{Nakamura2020,Guo2020a}, and threshold effects \cite{Rosner2007}.
	
A critical puzzle that motivates the present work is the anomalous ratio of the branching fractions for $Z_c(4430)$ (and the related structure). Experimentally, these states exhibit a strong preference for decaying into $\psi(2S)\pi$ over $J/\psi\pi$, with a ratio $R \approx 10$. This behavior stands in sharp contrast to the ground state $Z_c(3900)$, where the $J/\psi\pi$ is the discovery mode and the $\psi(2S)\pi$ mode is suppressed. This observation strongly suggests that these high-mass states possess an internal structure different from the $S$-wave molecules; specifically, they are likely composed of excited $P$-wave charmed mesons. The radially excited nature or the spatial extension due to the $P$-wave centrifugal barrier would naturally facilitate the overlap with the $\psi(2S)$ wave function, thereby explaining the enhanced decay width.

	However, investigating molecular states involving $P$-wave constituents---such as the scalar $D_0^*(2300)$ (denoted as $D_0^*$), the axial-vector $D_1(2420)/D_1(2430)$ (denoted as $D_1/D_1^\prime$), or the tensor $D_2^*(2460)$ (denoted as $D_2^*$)---requires moving beyond standard approximations that treat constituents as stable particles. Such simplified approaches neglect the significant decay widths of these mesons (primarily into $D^{(*)}\pi$) and the resulting three-body dynamics. Given their inherent instability, the molecular systems $D^*\bar{D}_1^\prime$, $D^*\bar{D}_1$, and $D^*\bar{D}_2^*$ are inevitably coupled to three-body continua.
	As demonstrated in previous studies on the $D\bar{D}^*$ system \cite{Cheng2022b,Lin2024b}, the three-body decay effects arising from the static limit correction (recoil effect) and self-energy diagrams can significantly shift the pole positions and, in particular, drastically amplify the widths. Given that the phase space and energy transfer for the decays of $P$-wave charmed mesons are much larger than those for $D^*$, these effects are expected to be even more pronounced in the present systems. Therefore, a systematic study is required. In this work, we investigate the $1^+(1^{+-})$ $Z_c$ systems composed of an $S$-wave and a $P$-wave charmed meson, explicitly incorporating the open-charm three-body decay effects and analyzing the corresponding line shapes.
	
	The paper is organized as follows. In Sec.~\ref{sec:framework}, we introduce the theoretical framework, including the coupled-channel formalism and the complex scaling method (CSM) incorporating width effects for unstable particles. In Sec.~\ref{sec: lagrangians and potential}, we construct the effective Lagrangians based on heavy quark and chiral symmetries, estimate the coupling constants, and derive the OBE potentials. In Sec.~\ref{sec: results}, we present the numerical results, analyzing the pole positions under the instantaneous approximation, static limit correction, and self-energy contributions. We also discuss the identification of the $Z_c(4430)$ and $Z_c(4200)$ candidates and analyze the line shape of the latter. Finally, a summary is given in Sec.~\ref{sec:summary}.
	
	%%%%%%%%%%%%%%%%%%%%%%\href{}{}%%%%%%%%%%%%%%%%%%%%%%%%%%%%%%%%

%%%%%%%%%%%%%%%%%%%%%%%%%%%%%%%%%%%%%%%%%%%%%%%%%%%%%%
	
	\section{Framework}\label{sec:framework}
	
In this work, we investigate the nature of $Z_c(4430)$ and $Z_c(4200)$ by considering interactions involving the $(D, D^*)$, $(D_0^*, D_1^\prime)$, and $(D_1, D_2^*)$ doublets with spin-parity $(0^-, 1^-)$, $(0^+, 1^+)$, and $(1^+, 2^+)$, respectively. To perform a systematic analysis of the $I^G(J^{PC})=1^+(1^{+-})$ hidden-charm sector, coupled-channel effects are explicitly included. This leads to 10 subsystems with 15 distinct channels, see Table \ref{tab: channels}. The first seven sectors are single-channel systems. The 8th ($D^*D_{1}^\prime$) and 9th ($D^*D_{1}$) sectors involve three partial waves ($^1P_1$, $^3P_1$, $^5P_1$), whereas the 10th sector ($D^*D_{2}^*$) involves two ($^3P_1$, $^5P_1$).

In previous works \cite{Cheng2022b,Lin2024a}, we investigated double- and hidden-charm tetraquarks using the CSM. We found that the unstable nature of the $D^*/\bar{D}^*$ mesons contributes an imaginary part to the OPE potential. This phenomenon is referred to as the three-body decay effect in our framework. Consequently, the $DD^*(^3S_1)$ bound state, which corresponds to the $T_{cc}^+$, acquires a small width. In this article, we assume that similar interactions exist in the OPE terms of the $D\bar{D}_0^* \leftrightarrow D_0^*\bar{D}$, $D^*\bar{D}_1^\prime \leftrightarrow D_1^\prime\bar{D}^*$, $D\bar{D}_2^* \leftrightarrow D_2^*\bar{D}$, $D^*\bar{D}_1 \leftrightarrow D_1\bar{D}^*$, and $D^*\bar{D}_2^* \leftrightarrow D_2^*\bar{D}^*$ channels.
	
For molecular states composed of these unstable particles, we must also consider the three-body decay effects arising from the self-energy terms of the unstable components, specifically $D_0^*$, $D_1^\prime$, $D_1$, and $D_2^*$. Based on our previous analysis \cite{Lin2024a}, the contribution of these self-energy terms to the total width is expected to be comparable to, or even larger than, that of the three-body effects in the OPE potential. Therefore, we will also investigate the impact of these self-energy effects on the $Z_c$ system.

%	这里我们并没有继续考虑$D\bar D^*\leftrightarrow D^*\bar D$ 过程中的三体衰变效应，这是因为其中$D^*$的宽度非常小。对于非常靠近阈值的奇特强子，如$T_{cc}^+$和 $X(3872)$，其三体衰变道是不可以忽略的。但在目前体系中，相较于$D^*$\to$ D\pi$衰变贡献的宽度，$D_0^*$, $D_1^{'}$, $D_1$ 和 $D_2^*$强子的衰变宽度要大得多。因此，在这个工作中，我们会将$D^*$强子也假设为一个没有宽度的稳定粒子，正如$D$介子一样。另一方面，因为这些粲介子与D或D^*介子的质量差要明显大于\pi介子，它们衰变时受到同位旋破缺的影响很小。所以我们在本工作中会考虑同位旋守恒的情况。We take the isospin average masses of the charmed meson and exchanged light mesons, and show them in Table \ref{tab: mass meson}.

\begin{table}[htbp]
	\centering
	\renewcommand{\arraystretch}{1.0}
		\setlength\tabcolsep{1.5pt}
	\begin{tabular}{cccccc}
		\hline\hline
		$I^G(J^{PC})$ & $1^-(0^{-+})$ & $0^+(0^{-+})$ & $0^+(0^{++})$ & $1^+(1^{--})$ & $0^-(1^{--})$  \\
		Meson & $\pi$ & $\eta$ & $\sigma$ & $\rho$ & $\omega$\\
		$M$ (MeV) & 138.04 & 547.86 & 500 & 775.26 & 782.66\\
		\hline
		$J^P$ & $0^-$ & $1^-$ & $0^+$ & $1^+$ & \\
		Meson & $D$ & $D^*$ & $D_0^*$ & $D_1^\prime$ &  \\
		$M$ (MeV) & 1867.25 & 2008.55 & $2343\pm10$ & $2412\pm9$ &   \\
		$\Gamma$ (MeV) & - & - & $229\pm16$ & $314\pm29$ &  \\
		\hline
		$J^P$ & $1^+$& $2^+$&&&\\
		Meson &$D_1$ &$D_2^*$ &&& \\
		$M$ (MeV)& $2422.1\pm0.6$&$2461.1\pm0.7$ &&&\\
		$\Gamma$ (MeV)&$31.3\pm1.9$ &$47.3\pm0.8$ &&& \\
		\hline\hline
	\end{tabular}
	\caption{The average masses (in units of MeV) of the charmed and light mesons, which are taken from Ref. \cite{ParticleDataGroup:2024cfk}.}
	\label{tab: mass meson}
\end{table}

\begin{table*}[htbp]
	\centering
	\renewcommand{\arraystretch}{1.4}
	\setlength\tabcolsep{3.5pt}
	\begin{tabular}{ccccccccc}
		\hline\hline
		Channel & 1 & 2 & 3 & 4 & 5 & 6 & 7 & \\
		System & $\{D\bar{D}^*\}(^3S_1)$ & $D^*\bar{D}^*(^3S_1)$ & $\{D\bar{D}_0^{*}\}(^1P_1)$ & $\{D\bar{D}_1^\prime\}(^3P_1)$ & $\{D\bar{D}_1\}(^3P_1)$ & $[D\bar{D}_2^*](^5P_1)$ & $[D^*\bar{D}_0^*](^3P_1)$ & \\
		\hline
		Channel & 8 & 9 & 10 & 11 & 12 & 13 & 14 & 15 \\
		System & $[D^*\bar{D}_1^\prime](^1P_1)$ & $\{D^*\bar{D}_1^\prime\}(^3P_1)$ & $[D^*\bar{D}_1^\prime](^5P_1)$ & $[D^*\bar{D}_1](^1P_1)$ & $\{D^*\bar{D}_1\}(^3P_1)$ & $[D^*\bar{D}_1](^5P_1)$ & $\{D^*\bar{D}_2^*\}(^3P_1)$ & $[D^*\bar{D}_2^*](^5P_1)$ \\
		\hline\hline
	\end{tabular}
	\caption{The channels of the hidden-charm tetraquark systems with $I^G(J^{PC})=1^{+}(1^{+-})$. We adopt the following shorthand notations for simplicity: $[A,\bar{B}]=\frac{1}{\sqrt{2}}(A\bar{B}-B\bar{A})$ and $\{A,\bar{B}\}=\frac{1}{\sqrt{2}}(A\bar{B}+B\bar{A})$.}
	\label{tab: channels}
\end{table*}

	%----------------------------------------------------------------------------------
	
	\subsection{A brief discussion on the CSM}\label{subsec:csm}
	
	In this work, we adopt the complex scaling method (CSM), originally proposed by Aguilar, Balslev, and Combes in the 1970s \cite{aguilarClassAnalyticPerturbations1971,balslevSpectralPropertiesManybody1971b}, along with the improved complex scaling method \cite{Chen2024}, to investigate bound states, resonances, and virtual states in momentum space. The transformations of the coordinate $r$ and momentum $k$ in the CSM are defined as:
	\begin{equation}
		U(\theta)r = r e^{i\theta}, \qquad U(\theta)k = k e^{-i\theta}. \label{eq:rktrans}
	\end{equation}

Following the complex scaling operation, the Schrödinger equation
\begin{equation}
	\frac{p^2}{2m}\phi_l(p)+ \int \frac{p^{\prime 2}dp^\prime}{(2\pi)^3} V_{l,l^\prime} (p,p^\prime)\phi_{l^\prime}(p^\prime)=E\phi_l(p), \label{MSE}
\end{equation}
in momentum space transforms into
\begin{align}
	&\frac{p^2e^{-2i\theta}}{2m}\tilde{\phi}_l(p)+ \int \frac{p^{\prime 2}e^{-3i\theta}dp^\prime}{(2\pi)^3} V_{l,l^\prime} (pe^{-i\theta},p^\prime e^{-i\theta})\tilde{\phi}_{l^\prime}(p^\prime) \nonumber\\
	&=E\tilde{\phi}_l(p), \label{eq:SECSM}
\end{align}
subject to the normalization condition \cite{Lin2023a}
\begin{eqnarray}
	&&\frac{e^{-3i\theta}}{(2\pi)^3}\int_{0}^{\infty}\tilde{\phi}_l(p)^2 p^2 dp=1,\label{eq:NM}
\end{eqnarray}
where $l$ and $l^\prime$ denote the orbital angular momenta, and $p$ represents the relative momentum in the center-of-mass frame. The potential $V_{l,l^\prime}$, obtained after partial wave decomposition, is expressed as
\begin{align}
	&V_{l,l^\prime} = \int d\Omega^\prime\int d\Omega\sum_{m_{l^\prime}=-l^\prime}^{l^\prime}\langle l^\prime,m_{l^\prime};s,m_j-m_{l^\prime}|j,m_j\rangle\nonumber\\
	&\qquad \times\sum_{m_{l}=-l}^{l}\langle l,m_{l};s,m_j-m_{l}|j,m_j\rangle \mathcal{Y}_{l^\prime,m_{l^\prime}}^*(\theta^\prime,\phi^\prime) \nonumber\\
	&\qquad \times \mathcal{Y}_{l,m_{l}}(\theta,\phi)\langle s,m_j-m_{l^\prime}|\mathcal{V}|s,m_j-m_l\rangle,\label{eq:Vpartial}
\end{align}
Here, $s$ and $j$ represent the total spin and total angular momentum of the system, respectively, and $m_l$ is the corresponding magnetic quantum number. $\mathcal{Y}_{l,m_{l}}(\theta,\phi)$ denotes the spherical harmonics associated with the angular coordinates $\theta$ and $\phi$. The potential operator $\mathcal{V}$ acts on the states $|s,m_j-m_{l^\prime}\rangle$ and $|s,m_j-m_l\rangle$.

We introduce a non-local monopole form factor to suppress potential contributions in the high-momentum region, defined as
\begin{align}
	F(p^{\prime 2}, p^2) &= \frac{\Lambda^2}{\Lambda^2+p^{\prime 2}}\frac{\Lambda^2}{\Lambda^2+p^2},
\end{align}
where $\Lambda$ represents the cutoff parameter.

 In the present work, we treat the $D$ and $D^*$ mesons as stable particles, whereas the broad states $D_0^*$, $D_1^\prime$, $D_1$, and $D_2^*$ are considered unstable. To incorporate the effects of their decays, we modify the complex-scaled Schrödinger equation in momentum space as follows:
\begin{align}
	&\left(\frac{p^2e^{-2i\theta}}{2m}-\frac{i}{2}\Gamma(E,pe^{-i\theta})\right)\tilde{\phi}_l(p) \nonumber\\
	&\qquad + \int \frac{p^{\prime 2}e^{-3i\theta}dp^\prime}{(2\pi)^3} V_{l,l^\prime} (pe^{-i\theta},p^\prime e^{-i\theta})\tilde{\phi}_{l^\prime}(p^\prime) = E\tilde{\phi}_l(p), \label{eq:SECSM2}
\end{align}
where $\Gamma(E, pe^{-i\theta})$ represents the width of the unstable meson following the complex scaling transformation. 

Taking the $DD_0^*$ system as an example, the expression for the width prior to complex scaling is given by
\begin{align}
	\Gamma_{D_0^*}(E,p)&=\frac{3}{2}\frac{f^{\prime\prime 2}}{2\pi f_\pi^2}\frac{m_D}{m_{D_0^*}} k_{\pi}E_\pi^2. \label{eq:width}
\end{align}
The factor of $3/2$ accounts for isospin, corresponding to the sum of the charged ($\pi^\pm$, factor $1$) and neutral ($\pi^0$, factor $1/2$) channels. Here, $k_{\pi}$ and $E_{\pi}$ denote the momentum and energy of the final-state pion, respectively defined as
\begin{align}
	k_{\pi}&=\frac{1}{2E_{D_0^*}}\lambda^{1/2}(E_{D_0^*}^2,m_D^2,m_\pi^2),\nonumber\\
	E_{D_0^*}&=[(E-\sqrt{p^2+m_D^2})^2-p^2]^{1/2}, \nonumber \\
	\lambda(a,b,c)&=a^2+b^2+c^2-2ab-2bc-2ac, \label{eq:width_sup}
\end{align}
where $E_{D_0^*}$ represents the effective invariant mass of the $D_0^*$ component when the total energy of the system is $E$ and the spectator $D$ meson is on-shell.
	
	\begin{figure}[htbp]
		\includegraphics[width=210pt]{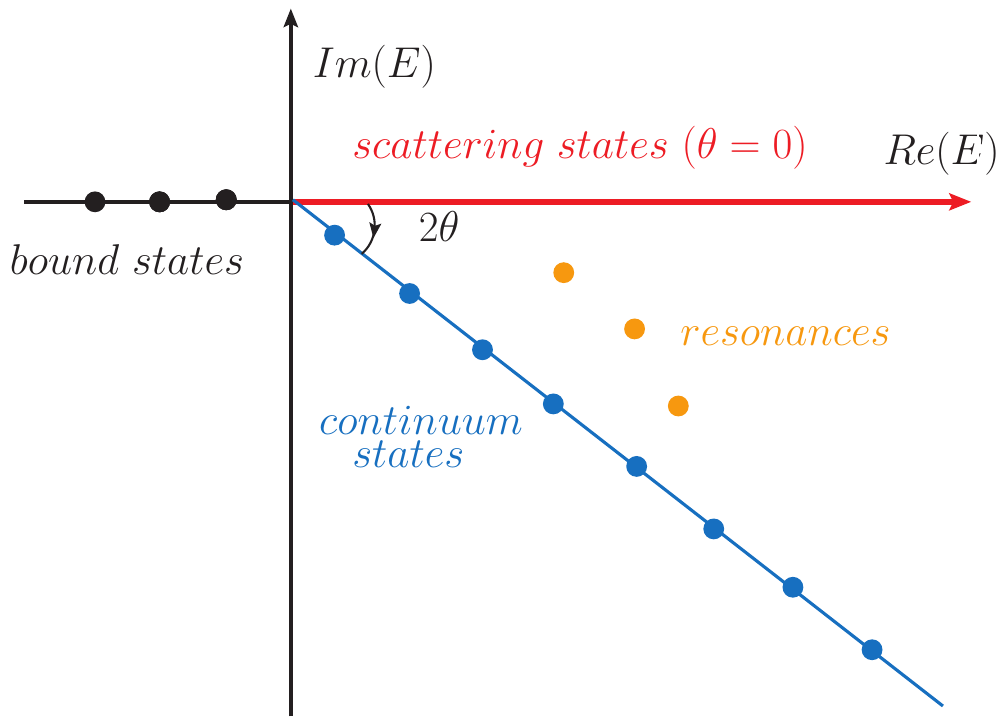}
		\caption{The eigenvalue distribution of the complex scaled Schr\"odinger equation for the two-body systems. }\label{fig: CSM plot}
	\end{figure}

\section{Lagrangians and Potentials}\label{sec: lagrangians and potential}

	\subsection{Lagrangians}\label{sec: lagrangians }	
	
	We adopt the OBE potential to describe the interactions between two heavy mesons. The effective Lagrangians are constructed based on heavy quark symmetry and chiral symmetry. The explicit expressions are given by
	\begin{align}
		&\mathcal{L}_{\pi/\eta} = g\langle H_b \slashed{A}_{ba}\gamma_5 \bar{H}_a\rangle + g^\prime\langle S_b \slashed{A}_{ba}\gamma_5 \bar{S}_a\rangle + g^{\prime\prime}\langle T_{\mu b} \slashed{A}_{ba}\gamma_5 \bar{T}_a^\mu\rangle \nonumber\\
		&\qquad + f^{\prime\prime}\langle S_b \slashed{A}_{ba}\gamma_5 \bar{H}_a\rangle + f^\prime\langle T_b^\mu A_{\mu ba}\gamma_5 \bar{S}_a\rangle \nonumber\\
		&\qquad + \frac{i}{\Lambda_\chi}\langle T_b^\mu(h_1 D_\mu \slashed{A}+h_2 \slashed{D} A_\mu)_{ba}\gamma_5\bar{H}_a\rangle + \text{H.c.},\\
		% ---
		&\mathcal{L}_{\rho/\omega} = i\beta\langle H_b v^\mu(V_\mu-\rho_\mu)_{ba}\bar{H}_a\rangle + i\lambda\langle H_b \sigma^{\mu\nu} F_{\mu\nu}(\rho)_{ba}\bar{H}_a\rangle \nonumber\\
		&\qquad + i\beta_1\langle S_b v^\mu(V_\mu-\rho_\mu)_{ba}\bar{S}_a\rangle + i\lambda_1\langle S_b \sigma^{\mu\nu} F_{\mu\nu}(\rho)_{ba}\bar{S}_a\rangle \nonumber\\
		&\qquad + i\beta_2\langle T_b^\mu v^\nu(V_\nu-\rho_\nu)_{ba}\bar{T}_{\mu a}\rangle + i\lambda_2\langle T_b^\mu \sigma^{\alpha\beta} F_{\alpha\beta}(\rho)_{ba}\bar{T}_{a\mu}\rangle \nonumber\\
		&\qquad + i\zeta\langle H_b \gamma^\mu(V_\mu-\rho_\mu)_{ba}\bar{S}_a\rangle + i\mu\langle H_b \sigma^{\mu\nu} F_{\mu\nu}(\rho)_{ba}\bar{S}_a\rangle \nonumber\\
		&\qquad + i\zeta_1\langle T_b^\mu (V_\mu-\rho_\mu)_{ba}\bar{H}_a\rangle + \mu_1\langle T_b^\mu \gamma^\nu F_{\mu\nu}(\rho)_{ba}\bar{H}_a\rangle \nonumber\\
		&\qquad + \mu_2\langle T_b^\mu \gamma^{\nu} F_{\mu\nu}(\rho)_{ba}\bar{S}_a\rangle+ \text{H.c.}, \\
		% ---
		&\mathcal{L}_{\sigma} = g_\sigma\langle H_a\sigma\bar{H}_a\rangle + g^\prime_\sigma\langle S_a\sigma\bar{S}_a\rangle + g^{\prime\prime}_\sigma\langle T_a^\mu\sigma\bar{T}_{a\mu}\rangle \nonumber\\
		&\qquad + \frac{ih_\sigma}{f_\pi}\langle S_a(\partial_\mu\sigma)\gamma^\mu\bar{H}_a\rangle + \frac{ih^\prime_\sigma}{f_\pi}\langle T_a^\mu(\partial_\mu\sigma)\bar{H}_a\rangle + \text{H.c.} \label{eq:lagrangian}
	\end{align}
Here, the superfields $H_a$, $S_a$, and $T_a^\mu$ represent the heavy meson doublets with spin-parity content $(0^-,1^-)$, $(0^+,1^+)$, and $(1^+,2^+)$, respectively. Their explicit expressions are given by
{\small\begin{align}
	H_a^{(Q)} &= \frac{1+\slashed{v}}{2}\left[P_a^{*\mu}\gamma_{\mu}-P_a\gamma_5\right], \label{eq:fieldH} \\
	S_a^{(Q)} &= \frac{1+\slashed{v}}{2}\left[P_{1a}^{\prime\mu}\gamma_{\mu}\gamma_5-P_{0a}^*\right], \label{eq:fieldS} \\
	T_a^\mu &= \scalebox{0.95}{\ensuremath{\displaystyle
			\frac{1+\slashed{v}}{2}\left\{P_{2a}^{*\mu\nu}\gamma_\nu-\sqrt{\frac{3}{2}}P_{1a}^\nu \gamma_5\left[g^\mu_\nu-\frac{1}{3}\gamma_\nu(\gamma^\mu-v^\mu)\right]\right\}
	}}, \label{eq:fieldT}
\end{align}}
where $a$ denotes the flavor index, and the multiplet components are defined as $P_a^{(*)}=\left(D^{(*)0},D^{(*)+},D_s^{(*)+}\right)$. Similarly, the components of the excited doublets are defined as $P_{0a}^*=\left(D_0^{*0},D_0^{*+},D_{s0}^{*+}\right)$, $P_{1a}^{\prime}=\left(D_1^{\prime0},D_1^{\prime+},D_{s1}^{\prime+}\right)$, $P_{1a}=\left(D_1^{0},D_1^{+},D_{s1}^{+}\right)$, and $P_{2a}^*=\left(D_2^{*0},D_2^{*+},D_{s2}^{*+}\right)$.

The fields involving light mesons are defined as follows:
\begin{align}
	\mathbb{A}_\mu &= \frac{i}{2} \left[\xi^\dagger(\partial_\mu\xi)+(\partial_\mu\xi)\xi^\dagger\right], \\
	V_\mu &= \frac{i}{2} \left[\xi^\dagger(\partial_\mu\xi)-(\partial_\mu\xi)\xi^\dagger\right],
\end{align}
with $\xi = \exp(i\mathcal{M}/f_\pi)$, where the pseudoscalar meson matrix is
\begin{equation}
	\mathcal{M} =
	\renewcommand{\arraystretch}{1.5}
	\begin{pmatrix}
		\frac{\pi^0}{\sqrt{2}}+\frac{\eta}{\sqrt{6}} & \pi^+ & K^+ \\
		\pi^- & -\frac{\pi^0}{\sqrt{2}}+\frac{\eta}{\sqrt{6}} & K^0 \\
		K^- & \bar{K}^0 & -\frac{2}{\sqrt{6}}\eta
	\end{pmatrix}.
\end{equation}
The vector meson field is given by $\rho^\mu = i \frac{g_V}{\sqrt{2}}\hat{\rho}^\mu$, with
\begin{equation}
	\hat{\rho}^\mu =
	\renewcommand{\arraystretch}{1.5}
	\begin{pmatrix}
		\frac{\rho^0}{\sqrt{2}}+\frac{\omega}{\sqrt{2}} & \rho^+ & K^{*+} \\
		\rho^- & -\frac{\rho^0}{\sqrt{2}}+\frac{\omega}{\sqrt{2}} & K^{*0} \\
		K^{*-} & \bar{K}^{*0} & \phi
	\end{pmatrix}^\mu.
\end{equation}
Here, the pion decay constant is taken as $f_\pi = 132$ MeV. The remaining coupling constants are determined via the quark model. The specific values adopted in our calculation are summarized in Table \ref{tab:couplings}. Detailed discussions regarding their extraction and the relevant coupling relations are provided in Appendix \ref{sec: couplings}.

\begin{table}[htbp]
	\caption{Coupling constants used in the calculation.}
	\label{tab:couplings}
	\centering
	\renewcommand{\arraystretch}{1.3}
	\begin{tabular}{cccccc}
		\hline\hline
		$g$ & $g^{\prime}$ & $g^{\prime\prime}$ & $f^{\prime}$ & $f^{\prime\prime}$ & $\frac{h_1+h_2}{\Lambda_\chi}$   \\
		$-0.59$ & $0.25$ & $0.75$ & $-0.87$ & $0.53$ & $1.03~\text{GeV}^{-1}$  \\
		\hline
		$\beta$ & $\beta_1$ & $\beta_2$ & $\lambda$ & $\lambda_1$ & $\lambda_2$ \\
		$0.9$ & $-0.9$ & $-0.9$ & $0.56~\text{GeV}^{-1}$ & $0.19~\text{GeV}^{-1}$ & $-0.56~\text{GeV}^{-1}$\\
		$g_V$ & $\zeta$ & $\zeta_1$ &  $\mu$ & $\mu_1$ & $\mu_2$ \\
		$5.8$ & $0.81$ & $1.40$ &  $-1.00~\text{GeV}^{-1}$ & $-1.74~\text{GeV}^{-1}$ & $-1.29~\text{GeV}^{-1}$\\
		\hline
		$g_\sigma$&$g_\sigma^{\prime}$&$g_\sigma^{\prime\prime}$&$h_\sigma$&$h_\sigma^{\prime}$&\\
		2.73&-2.73&-2.73&0.48&-0.48&\\
		\hline\hline
	\end{tabular}
\end{table}

The explicit expressions for the expanded Lagrangians describing the interactions between heavy flavor mesons and light mesons $(\pi,\sigma,\rho/\omega)$ are detailed in Appendix \ref{appendix:lagrangian}.

\subsection{Potential}\label{sec: potentials}

The explicit expressions for the OBE potentials in the diagonal sectors are presented in momentum space in Appendix~\ref{appendix:potential}. As an illustrative example, consider the $\{D\bar D^*\}(^3S_1)$ channel:
\begin{align}
	V(D&\bar D^*\to D\bar D^*) = \delta_C\frac{g^2}{2f_\pi^2}(\vec q \cdot \vec \epsilon_3^\dagger)(\vec q \cdot \vec \epsilon_1) \nonumber \\
	&\times \left(\frac{\vec\tau_1\cdot\vec \tau_2}{q^2-m_\pi^2}-\frac{1/3}{q^2-m_\eta^2}\right) +g_\sigma^2 \frac{\vec\epsilon_4^\dagger\cdot \vec\epsilon_2}{q^2-m_\sigma^2} \nonumber\\
	&+ g_V^2 \bigg\{-\frac{1}{4}\beta^2 (\vec\epsilon_4^\dagger\cdot \vec\epsilon_2)-\delta_C\lambda^2 \Big[\vec q^2(\vec\epsilon_3^\dagger\cdot \vec\epsilon_2) \nonumber\\
	&-(\vec q\cdot \vec\epsilon_3^\dagger)(\vec q\cdot \vec\epsilon_2)\Big]\bigg\} \left(\frac{\vec{\tau}_1\cdot\vec{\tau}_2}{q^2-m_\rho^2}-\frac{1}{q^2-m_\omega^2}\right).
\end{align}
where $\vec\epsilon$ denotes the polarization vector of the $D^*/\bar{D}^*$ meson. The indices $1,2,3,4$ label the particles, such that particles $1$ and $3$ ($2$ and $4$) share a vertex. $\delta_C$ represents a phase factor arising from the cross-channel diagrams, determined by the symmetry of the final state flavor configuration: $\delta_C=1$ for $\{A\bar B\}$ and $\delta_C=-1$ for $[A\bar B]$. Obviously, $\delta_C=1$ for the process $\{D\bar D^*\}(^3S_1)\to\{D\bar D^*\}(^3S_1)$.

As discussed in Refs. \cite{Cheng2022b,Cheng2023b,Lin2024b}, in the OPE cross-channel diagrams, the condition $|q^0| \approx m_{D^*} - m_D > m_\pi$ can induce a three-body mediated decay effect. This effect significantly influences the imaginary part of the pole near the threshold. It becomes particularly pronounced for the $D^{(*)}\bar{D}_0^*$, $D^{(*)}\bar{D}_1^\prime$, $D^{(*)}\bar{D}_1$, and $D^{(*)}\bar{D}_2^*$ cases, where the energy transfer $|q^0|$ is much larger than that in the $D\bar{D}^*$ case.

Assuming the final state is at rest, the transferred energy carried by the exchanged light meson is given by
\begin{equation}
	q^0=\frac{m_1^2-m_2^2-m_3^2+m_4^2}{2(m_3+m_4)},
\end{equation}
where indices $1,2$ ($3,4$) correspond to the initial (final) particles. In the diagonal sector, this static limit approximation implies zero recoil correction, as the initial and final kinetic energies both vanish. We list the $q^0$ values for the direct ($q_{0D}$) and cross ($q_{0C}$) diagrams in the diagonal channels in Table \ref{tab: q0}. It is evident that $q_{0D}=0$ for direct diagrams, while $q_{0C} = m_1 - m_2$. When $|q^0| > m_\pi$, the three-body decay effect emerges, potentially affecting the particle widths significantly.

\begin{table}[htbp]
	\centering
	\renewcommand{\arraystretch}{1.0}
	\setlength\tabcolsep{8pt}
	\begin{tabular}{cccccc}
		\hline\hline
		System & $D\bar D^*$ & $D^*\bar D^*$ & $D\bar D_0^{*}$ & $D\bar D_1^\prime$ & $D\bar D_1$\\
		$q_{0D}$ & 0 & 0 & 0 & 0 & 0\\
		$q_{0C}$ & $-141$ & 0 & $-476$ & $-545$ & $-555$\\
		\hline
		System & $D\bar D_2^*$ & $D^*\bar D_0^*$ & $D^*\bar D_1^\prime$ & $D^*\bar D_1$ & $D^*\bar D_2^*$\\
		$q_{0D}$ & 0 & 0 & 0 & 0 & 0\\
		$q_{0C}$ & $-594$ & $-334$ & $-403$ & $-414$ & $-453$\\
		\hline\hline
	\end{tabular}
	\caption{The transferred energy values $q_{0D}$ and $q_{0C}$ (in units of MeV) for the diagonal channels calculated in the static limit. Note that $q_{0D}$ vanishes for all diagonal channels in this approximation.}
	\label{tab: q0}
\end{table}

In the isovector molecular case, it is straightforward to observe that the One-Vector-Exchange (OVE) term is proportional to $(1/(q^2-m_\rho^2) - 1/(q^2-m_\omega^2))$, where the two terms largely cancel each other. Consequently, the OVE contribution is negligible. Similarly, the $\eta$ exchange term is suppressed to about $1/3$ of the pion strength. Therefore, the dominant interactions arise from the OPE and One-Sigma-Exchange (OSE) potentials.
At this stage, the determination of $g_\sigma$ becomes crucial. In Ref. \cite{Zhu:2024hgm}, the authors utilized the specific exotic state $Z_c(3900)$ as input to constrain $g_\sigma$. In the present study, as detailed in the Appendix~\ref{sec: couplings}., we adopt the chiral quark model, setting $g_\sigma=2.73$.

\begin{figure*}[htbp]
	\raggedright
	\subfigure{ 
		\includegraphics[width=500pt]{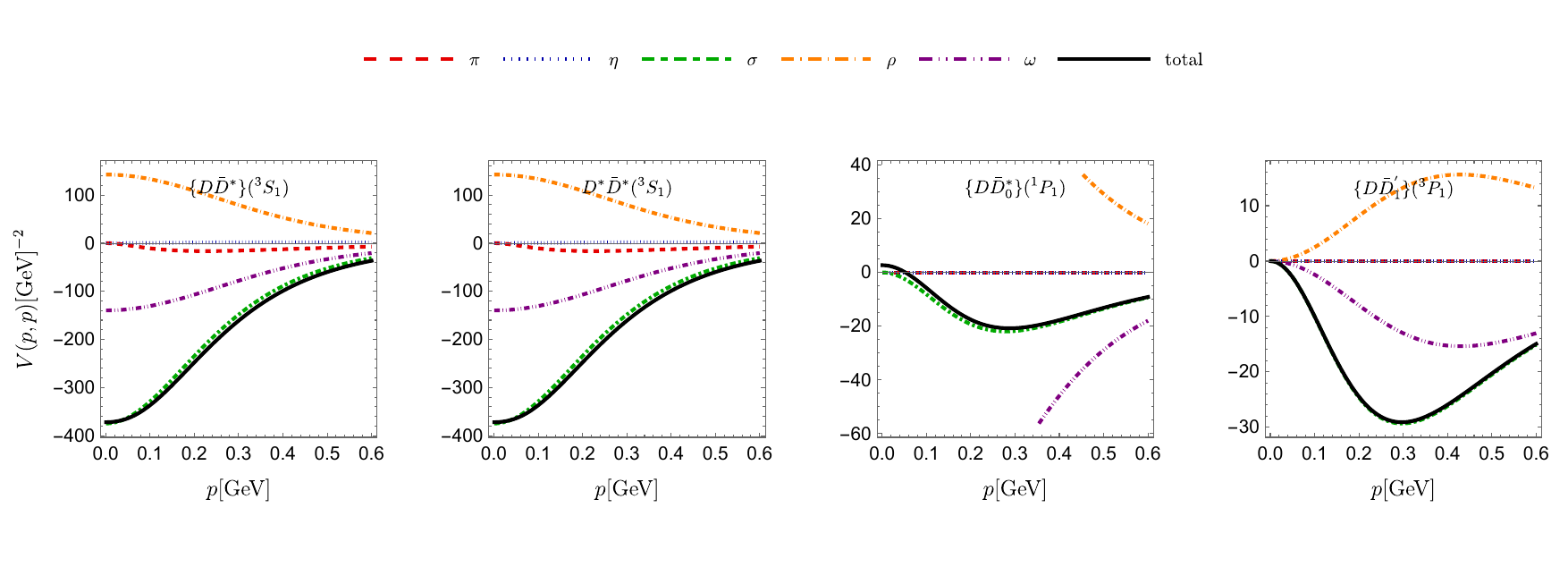}}
	\par\vspace{-20pt}
	\subfigure{ 
		\includegraphics[width=380pt]{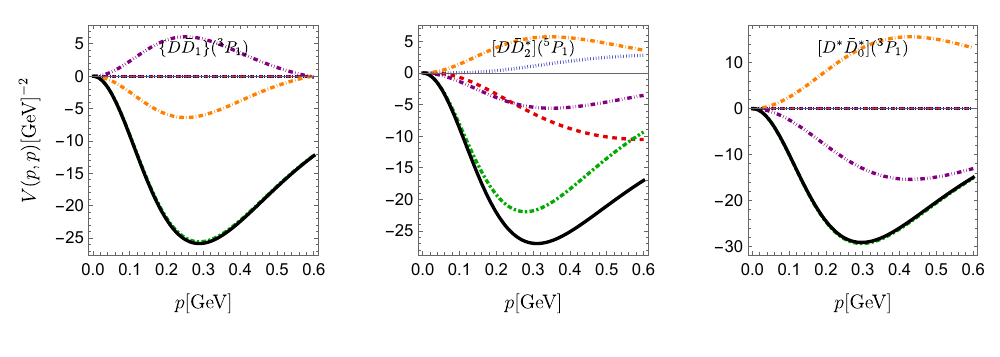}}
	\caption{\small The OBE potentials for the diagonal channels:
		$\{D\bar D^*\}(^3S_1)$, $D^*\bar D^*(^3S_1)$, $\{D\bar D_0^{*}\}(^1P_1)$, $\{D\bar D_1^\prime\}(^3P_1)$, $\{D\bar D_1\}(^3P_1)$, $[D\bar D_2^*](^5P_1)$, and $[D^*\bar D_0^*](^3P_1)$,
		with quantum numbers $I^G(J^{PC})=1^+(1^{+-})$. The results are obtained using the instantaneous approximation ($q_0=0$) and a non-local regulator with cutoff $\Lambda=0.6$ GeV.}\label{fig: potential}
\end{figure*}

The partial-wave projected potentials for the diagonal sectors are also provided in Appendix~\ref{appendix:potential}. Notably, for the spin-mixing transition $(^1P_1) \to (^3P_1)$ in the $D^*D_1^\prime$ sector, the potential vanishes ($V_{89}=0$). However, this is not the case for the $(^1P_1) \to (^5P_1)$ transition. This phenomenon was also observed in a recent study of the $\bar D^* K^*$ system \cite{Wang:2024ukc}, where the potential for the $(^{1,5}P_1) \to (^3P_1)$ process was found to be zero. The underlying reason is that both $\bar D^*$ and $K^*$ (or our $D^*$ and $D_1^\prime$) carry spin 1, and their spin wave functions possess specific symmetry properties under particle exchange. The $(^1P_1)$ and $(^5P_1)$ states share the same spin symmetry, whereas the $(^3P_1)$ state has opposite symmetry. Consequently, the matrix element of the potential between these states vanishes:
\begin{equation}
	\langle ^{3}P_1 | \hat{V} | ^{1,5}P_1 \rangle \equiv 0.
\end{equation}
Finally, we plot the diagonal sector potentials in Fig.~\ref{fig: potential}, and verify that a similar vanishing potential occurs in the $D^*D_1^\prime$ and $D^*D_1$ sectors.

\begin{figure*}[htbp]
	\raggedright
	\subfigure{ 
		\includegraphics[width=500pt]{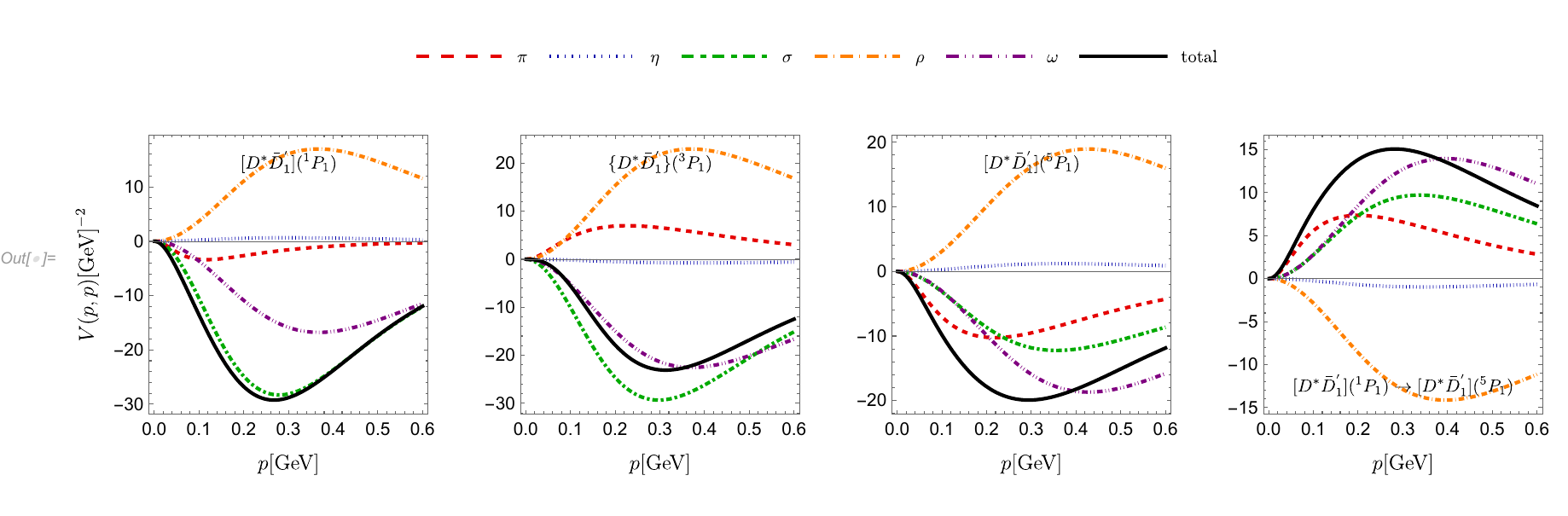}}
	\par\vspace{-20pt}
	\subfigure{ 
		\includegraphics[width=500pt]{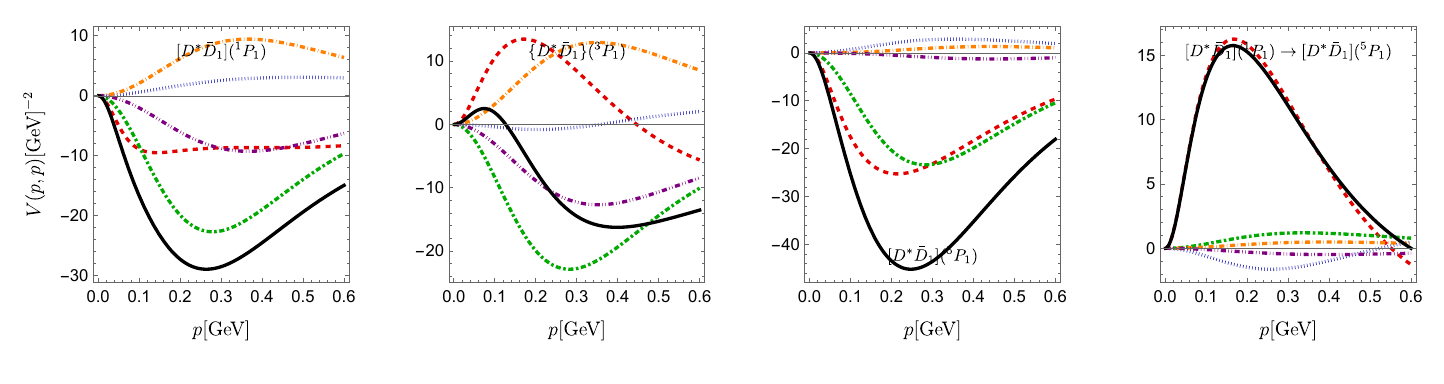}}
	\par\vspace{-20pt}
	\subfigure{ 
	\includegraphics[width=380pt]{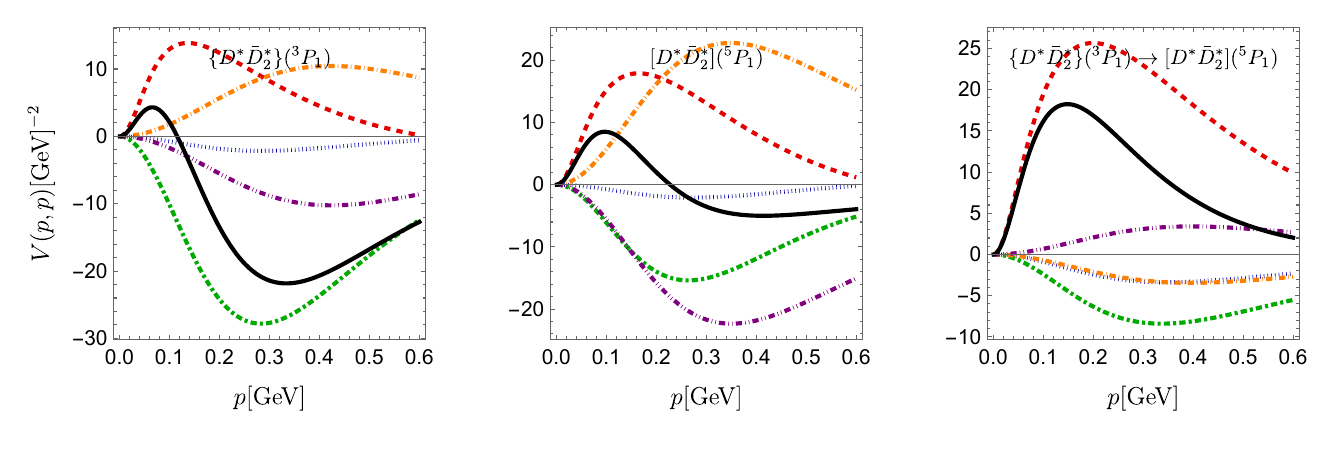}}	
	\caption{\small The OBE potentials for the diagonal sectors of
		$D^*\bar D_1^\prime$, $D^*\bar D_1$, and $D^*\bar D_2^*$
		with quantum numbers $I^G(J^{PC})=1^+(1^{+-})$. The potentials for the non-vanishing spin-mixing transitions among the $(^{1,3,5}P_1)$ partial waves are also displayed. The parameters are the same as in Fig.~\ref{fig: potential}.}\label{fig: potential2}
\end{figure*}

\section{Numerical results}\label{sec: results}

In our previous studies \cite{Lin2024a}, we interpreted the $T_{cc}^+$, $X(3872)$, $G(3900)$, and $Z_c(3900)$ as $DD^*/D\bar{D}^*$ molecules with quantum numbers $0(1^+)$, $0^+(1^{++})$, $0^-(1^{--})$, and $1^+(1^{+-})$, respectively.
In the present work, however, we introduce a different coupling constant $g_\sigma=2.73$ (compared to $g_\sigma=0.76$ in Ref. \cite{Lin2024a}) based on the chiral quark model estimation.
Since this larger coupling significantly enhances the attractive strength of the $\sigma$ exchange potential, the cutoff $\Lambda$ must be adjusted accordingly to maintain physical consistency.

\begin{table*}[htbp]
	\centering
	\renewcommand{\arraystretch}{1.3}
	\setlength\tabcolsep{8pt}
	\begin{tabular}{cccccc}
		\hline\hline
		System & $DD^*(^3S_1)$ & $[D\bar D^*](^3S_1)$ & $[D\bar D^*](^3P_1)$ & $[D\bar D^*](^3S_1)$ & $D^*\bar D^*(^3S_1)$ \\
		$I(J^{P}) / I^G(J^{PC})$ & $0(1^+)$ & $0^+(1^{++})$ & $0^-(1^{--})$ & $1^+(1^{+-})$ & $1^+(1^{+-})$ \\
		\hline
		$E$ ($g_\sigma=0.76$, $\Lambda=500$ MeV) & $-0.57^B$ & $-3.2^B$ & $-3.8-27.0i$ & $-25.8^V$ & $-24.4^V$ \\
		$E$ ($g_\sigma=2.73$, $\Lambda=350$ MeV) & $-1.3^B$ & $-3.8^B$ & $-7.4-18.6i$ & $-1.2^V$ & $-0.99^V$ \\
		\hline\hline
	\end{tabular}
	\caption{The eigenenergies (in units of MeV) relative to the corresponding thresholds: $3875.8$ MeV for the first four systems and $4017.1$ MeV for the fifth system. The extracted poles correspond to the molecular candidates for the exotic states $T_{cc}^+$, $X(3872)$, $G(3900)$, $Z_c(3900)$, and $Z_c(4020)$, respectively. The superscripts B and V denote bound states and virtual states, respectively; otherwise, the pole represents a resonance. In both schemes, the instantaneous approximation $q^0=0$ is adopted.}
	\label{tab: Tcc}
\end{table*}

Under the instantaneous approximation ($q^0=0$), we reconsider the $DD^*/D\bar{D}^*$ molecules and set the cutoff to $\Lambda=0.5$ GeV for the $g_\sigma=0.76$ scheme, while determining $\Lambda=0.35$ GeV for the $g_\sigma=2.73$ scheme.
The calculated results for the exotic states $T_{cc}^+$, $X(3872)$, $G(3900)$, $Z_c(3900)$, and $Z_c(4020)$ under these two parameter schemes are presented in Table~\ref{tab: Tcc}. 
In the $g_\sigma=2.73$ scheme, the $Z_c(3900)$ and $Z_c(4020)$ are identified as virtual states lying very close to their respective thresholds, which reflects a dominant molecular nature composed of $D^{(*)}\bar{D}^*$ components.
Despite this quantitative shift, we observe that the overall conclusions remain consistent with those reported in Ref. \cite{Lin2024a}.

However, the cutoff parameter determined from the $DD^*$ system may not be suitable for the sectors involving the $P$-wave $(D_0^*, D_1^\prime)$ and $(D_1, D_2^*)$ doublets. This difference in intrinsic parity leads to a potential structure distinct from the $S$-wave case.
Furthermore, as discussed in Appendix \ref{sec: couplings}, the momentum transfer $|\vec{q}|$ in the cross-channel diagrams is significantly higher than that in the $DD^*$ system, suggesting that a larger cutoff $\Lambda$ should be selected.
Consequently, we will determine the cutoff parameter for these sectors independently.

\subsection{The $P$-wave $Z_c$ systems}\label{Zc1}

We first present the results obtained under the instantaneous approximation ($q^0=0$), as listed in the third column of Table \ref{tab: results}. In the single-channel systems $\{D\bar D_0^{*}\}(^1P_1)$, $\{D\bar D_1^\prime\}(^3P_1)$, $\{D\bar D_1\}(^3P_1)$, $[D\bar D_2^*](^5P_1)$, and $[D^*\bar D_0^*](^3P_1)$, we consistently observe a similar resonance pole located below the threshold. This finding mirrors the analysis of the $G(3900)$ as a $P$-wave $D\bar D^*$ resonance in Ref. \cite{Lin2024a}, where we proposed that the centrifugal barrier inherent to $P$-wave systems facilitates the formation of resonant states. In contrast, for the $S$-wave $Z_c(3900)$ candidate, the attractive potential is deeper, favoring the formation of bound or virtual states. 
Due to the centrifugal barrier present in these $P$-wave systems, the wave function is effectively shielded from the short-range region. Consequently, the contributions from short-range interactions (such as $\eta$, $\rho$, and $\omega$ exchanges) are significantly reduced, rendering the system more sensitive to the medium- and long-range potentials dominated by $\pi$ and $\sigma$ exchanges. 

\begin{table*}[htbp]
	\centering
	\renewcommand{\arraystretch}{1.3}
	\setlength\tabcolsep{5pt}
	\begin{tabular}{cccccc}
		\hline\hline
		System & Threshold & & $\left[E,\frac{\mu k_{R,i}}{4\pi^2}|\text{Res}T|\right]_{\text{Inst}}$ & $\left[E,\frac{\mu k_{R,i}}{4\pi^2}|\text{Res}T|\right]_{\text{SL}}$ & $\left[E,\frac{\mu k_{R,i}}{4\pi^2}|\text{Res}T|\right]_{\text{SE}}$ \\
		\hline	
		$\{D\bar D_0^{*}\}(^1P_1)$ & 4210.3 & & $[-25.9-31.7i, (20.4)]$ & $[-21.6-26.1i, (19.1)]$ & $[8.7-184.2i, (45.7)]$ \\
		$\{D\bar D_1^\prime\}(^3P_1)$ & 4279.3 & & $[-21.7-33.5i, (22.4)]$ & $[-21.6-26.3i, (17.3)]$ & $[3.4-131.0i, (34.3)]$ \\
		$\{D\bar D_1\}(^3P_1)$ & 4289.4 & & $[-23.1-32.4i, (21.8)]$ & $[-20.6-27.8i, (17.7)]$ & $[-15.0-42.4i, (16.3)]$ \\
		$[D\bar D_2^*](^5P_1)$ & 4328.4 & & $[-24.0-32.8i, (23.0)]$ & $[-23.8-30.8i, (21.1)]$ & $[-9.8-78.0i, (25.6)]$ \\
		$[D^*\bar D_0^*](^3P_1)$ & 4351.6 & & $[-20.4-32.6i, (21.9)]$ & $[-18.5-31.5i, (21.4)]$ & $[17.6-195.1i, (51.3)]$ \\ \addlinespace
		
		\multirow{2}{*}{$[D^*\bar D_1^\prime](^1P_1,^5P_1)$} & \multirow{2}{*}{$4420.6$} & \ldelim\{{2}{3mm}[] & $[-16.5-29.6i, (17.3, 3.3)]$ & $[-11.8-21.2i, (12.4, 7.8)]$ & $[13.7-128.9i, (30.4, 17.1)]$ \\
		& & & $[-51.0-31.6i, (3.4, 25.3)]$ & $[-34.1-20.8i, (5.7, 18.7)]$ & $[-19.7-126.0i, (11.2, 32.9)]$ \\ \addlinespace
		
		$\{D^*\bar D_1^\prime\}(^3P_1)$ & 4420.6 & & $[-22.3-33.6i, (21.7)]$ & $[-23.1-34.2i, (21.7)]$ & $[3.0-141.6i, (36.4)]$ \\ \addlinespace
		
		\multirow{2}{*}{$[D^*\bar D_1(^1P_1,^5P_1)]$} & \multirow{2}{*}{$4430.7$} & \ldelim\{{2}{3mm}[] & $[-25.5-32.4i, (18.3, 5.1)]$ & $[-25.1-30.7i, (18.0, 4.3)]$ & $[-20.2-45.0i, (27.5,3.9)]$ \\
		& & & $[-15.1-22.9i, (4.2, 14.4)]$ & $[-14.7-23.1i, (4.1, 15.3)]$ & $[-9.4-37.9i, (3.9,16.1)]$ \\ \addlinespace
		
		$\{D^*\bar D_1\}(^3P_1)$ & 4430.7 & & $[-28.6-36.3i, (25.5)]$ & $[-27.6-34.0i, (23.9)]$ & $[-22.5-48.4i, (22.2)]$ \\ \addlinespace
		
		\multirow{2}{*}{$D^*\bar D_2^*(^3P_1,^5P_1)$} & \multirow{2}{*}{$4469.7$} & \ldelim\{{2}{3mm}[] & $[-24.8-37.1i, (17.8, 19.0)]$ & $[-21.6-29.7i, (21.2, 1.5)]$ & $\bm{[-8.1-76.1i, (26.5, 2.0)]}_{\textbf{A}}$ \\
		& & & $[-23.4-28.5i, (15.5, 15.9)]$ & $[-30.8-35.3i, (1.5, 21.5)]$ & $[-18.7-81.2i, (1.6, 22.6)]$ \\ \addlinespace	
		\hline\hline
	\end{tabular}
	\caption{The extracted poles of the $P$-wave systems with $I^G(J^{PC})=1^+(1^{+-})$ using a cutoff $\Lambda=0.6$ GeV. The values $\left[E,\frac{\mu k_{R,i}}{4\pi^2}|\text{Res}T|\right]$ represent the pole energy $E$ and the residue of the $T$-matrix, respectively. The data (in units of MeV) in the column ``Inst" are obtained under the instantaneous approximation ($q_0=0$). The column ``SL" refers to the static limit using the $q_0$ values from Table~\ref{tab: q0}. The column ``SE" further includes the self-energy contributions based on the static limit settings.}
	\label{tab: results}
\end{table*}

An examination of the potential forms in Eq.~\eqref{eq:V44}, \eqref{eq:V55}, \eqref{eq:V77} reveals that the $\{D\bar D_1^\prime\}(^3P_1)$, $\{D\bar D_1\}(^3P_1)$, and $[D^*\bar D_0^{*}](^3P_1)$ channels receive no contribution from the OPE term; their interactions are governed almost entirely by the OSE potential. Although the $\{D\bar D_0^{*}\}(^1P_1)$ channel contains an OPE term, its contribution is proportional to the transferred energy $q_{0C}$ in the cross-diagram, which vanishes in the instantaneous approximation. The $[D\bar D_2^*](^5P_1)$ channel does include an OPE contribution, but as shown in Fig.~\ref{fig: potential}, its magnitude is small. Therefore, considering Heavy Quark Spin Symmetry (HQSS), it is natural to observe highly similar pole positions and behaviors across these systems.

In the $D^*\bar D_1^\prime$ sector, spin-mixing occurs between the $(^1P_1)$ and $(^5P_1)$ partial waves, whereas the $(^3P_1)$ channel remains decoupled. Evidently, the behavior of the isolated $(^3P_1)$ channel parallels that of the other single-channel systems. The coupling strength between a pole and the $(^1P_1)$ or $(^5P_1)$ channel is reflected in the residue of the $T$-matrix. To quantify the effective coupling constant $g_Z$ via the residue, we introduce the relation:
\begin{equation}
	g_Z^2 \propto \lim_{E\to E_R}\frac{\mu k_0}{4\pi^2}(E-E_R)T_{ii}(E)=\left|\frac{\mu_j k_{R,i}}{4\pi^2}\langle k_{R,i}|\hat{V}|\phi\rangle^2\right|.\label{eq: 2}
\end{equation}
Here, $E_R$ and $k_{R,i}$ denote the energy and momentum of the pole in the $i$-th channel. In this work, we extract the residue of the $T$-matrix using the CSM and verify the results using the Complex Scaled Lippmann-Schwinger Equation (CSLSE) \cite{Wang2024}.

In the CSM framework, while the ratios between partial widths for broad resonances are accurate, the absolute magnitudes may not be precise \cite{Masui1999}. Typically, the partial widths (or residues) are determined by combining the total width (from the pole's imaginary part) with the calculated branching ratios. Based on the residue ratios, distinct coupling patterns emerge for the two poles found in the coupled $D^*\bar D_1^\prime(^{1,5}P_1)$ sector. The first pole couples predominantly to the $^{1}P_1$ channel, while the second couples mainly to the $^{5}P_1$ channel. We also calculated the wave function probabilities for these poles, obtaining $(88.2+6.4i, 11.8-6.4i)\%$ for the first pole and $(15.2-1.3i, 84.8+1.3i)\%$ for the second. These component probabilities are consistent with the residue analysis.

The pole positions and behaviors in the $D^*\bar D_1$ sector are largely similar to those in the $D^*\bar D_1^\prime$ sector. In the $D^*\bar D_2^*$ sector, two poles also emerge; however, they are closer in energy, and the residue ratios between their respective coupled channels are approximately 1:1, indicating strong mixing.

\subsection{The static limit and self-energy diagram correction}\label{Zc2}

We now proceed to incorporate three-body effects by considering the cross-diagrams in the OPE potential and the self-energy contributions of the constituent mesons.

The results calculated under the static limit approximation are presented in the fourth column of Table \ref{tab: results}. Compared to the instantaneous approximation scheme, both the eigenenergies and residues show observable shifts. To isolate the impact of the OPE-induced three-body effect, we use the $\{D\bar D_1^\prime\}(^3P_1)$, $\{D\bar D_1\}(^3P_1)$, and $[D^*\bar D_0^*](^3P_1)$ diagonal potentials as baselines; these channels lack OPE contributions and thus are free from OPE-mediated three-body decay. In contrast, the $\{D\bar D_0^{*}\}(^1P_1)$ and $[D\bar D_2^*](^5P_1)$ channels contain OPE terms with significant energy transfer. However, the variations in their pole positions and residues are comparable to those in the non-OPE channels. This suggests that while the static limit correction (manifested as non-zero $q_0$) does influence the pole behavior, the specific contribution from the OPE-induced three-body decay is relatively minor.

We further examine the influence of the static limit approximation on sectors with spin-mixing. For the $[D^*\bar D_1(^1P_1,^5P_1)]$ system, the pole behavior remains largely unchanged. In contrast, the $[D^*\bar D_1^\prime](^1P_1,^5P_1)$ system exhibits more noticeable shifts in energy and residue ratios. In the $D^*\bar D_2^*(^3P_1,^5P_1)$ system, while the two poles separate only slightly, their residues exhibit a remarkable change. Combining this with the calculated component probabilities---$(105.5-5.2i, -5.5+5.2i)\%$ for the first pole and $(-5.5+5.3i, 105.5-5.3i)\%$ for the second---we conclude that the first (second) pole has effectively decoupled from the $^5P_1$ ($^3P_1$) channel in the static limit approximation.

\begin{table*}[htbp]
	\centering
	\renewcommand{\arraystretch}{1.3}
	\setlength\tabcolsep{3pt}
	\begin{tabular}{ccccc}
		\hline\hline
		Pole & $E$ (MeV) & & Properties (\%) & $\frac{\mu k_{R,i}}{4\pi^2}|\text{Res}T|$ (MeV) \\
		\hline	
		% Pole A (Zc4430 candidate)
		\multirow{3}{*}{Pole A} & \multirow{3}{*}{$4461.0-75.7i$} & \ldelim\{{3}{3mm}[] &
		$(0, -0.1-0.1i, 0, 0, 0,$ & $(0.4, 0.2, 0, 4.3, 0.2,$ \\
		
		& & & $0, 0.1i, 0.2+0.8i, 0.3, 0.2-0.2i,$ & $0.7, 0.8, 0.6, 2.6, 1.5,$ \\
		
		& & & $-0.6i, 0.1i, -0.2i, \mathbf{115.2-18.8i}, \mathbf{-15.8+18.9i})$ & $0.1, 0.2, 0.1, \mathbf{28.4}, \mathbf{5.9})$ \\ 
		\hline % 分隔线
		
		% D0* Pole (Zc4200 candidate)
		\multirow{3}{*}{$\{D\bar D_0^*\}$ Pole} & \multirow{3}{*}{$4211.0-177.9i$} & \ldelim\{{3}{3mm}[] & 
		$(-0.1-0.2i, 0.8+0.2i, \mathbf{64.6+16.7i}, -1.8-0.3i, -0.3-0.3i,$ & $(1.7, 2.0, \mathbf{203.0}, 1.1, 1.0,$ \\
		
		& & & $-0.2i, -1.6+0.3i, 2.0+0.8i, -0.1+0.5i, 5.9+0.5i,$ & $0.5, 0.1, 0.7, 0.1, 1.2,$ \\
		
		& & & $22.4-25.0i, 3.5+8.6i, 6.1-5.4i, -1.2+4.5i, -0.2-0.8i)$ & $7.9, 1.5, 3.3, 2.0, 0.8)$ \\ 
		\hline\hline
	\end{tabular}
	\caption{The eigenenergies, component probabilities (Properties), and $T$-matrix residues of the representative candidate poles calculated in the full 15 coupled-channel space. ``Pole A'' corresponds to the $Z_c(4430)$ candidate identified in the $D^*\bar{D}_2^*$ sector (see the bolded entry in Table \ref{tab: results}), while ``$\{D\bar D_0^*\}$ Pole'' corresponds to the $Z_c(4200)$ candidate in the $\{D\bar{D}_0^*\}$ sector (see the first row of Table \ref{tab: results}). The dominant components are highlighted in bold.}
	\label{tab: mutichannel}
\end{table*}

Building upon the static limit, we next incorporate the self-energy effects of the unstable mesons $D_0^*$, $D_1^\prime$, $D_1$, and $D_2^*$. As indicated in Eq. \eqref{eq:SECSM2}, we explicitly include the imaginary part of the self-energy, treating the real part as a perturbation. The decay width formulas for these charmed mesons are provided in Appendix \ref{appendix:width}, assuming that the widths are dominated by two-body strong decays involving a pion. The resulting pole parameters are listed in the fifth column (``SE'') of Table \ref{tab: results}.

Consider the pole in the $\{D\bar D_0^{*}\}(^1P_1)$ channel. Its energy shifts dramatically compared to the previous two schemes: the real part moves from below to above the threshold, and the imaginary part increases significantly, far exceeding the value without self-energy corrections.
Clearly, there is a positive correlation between the large width of the constituent $D_0^*$ ($229$ MeV) and the final width of the molecular state.
A similar pattern is observed in other systems composed of broad mesons like $D_1^\prime$: the widths are amplified drastically, resulting in broad near-threshold resonances.
Given the large experimental uncertainties of the $Z_c(4200)$, the broad resonances in both the $\{D\bar D_0^{*}\}(^1P_1)$ and $\{D\bar D_1^\prime\}(^3P_1)$ sectors fall within a reasonable range to be considered as candidates.

For systems containing the relatively narrower $D_1$ and $D_2^*$, the real parts of the poles move closer to the threshold but remain below it, while the imaginary parts increase correspondingly with the widths of these constituents.
Notably, the $D^*\bar D_2^*(^3P_1,^5P_1)$ and $D^*\bar D_1(^{1,3,5}P_1)$ sectors host multiple broad poles with masses and widths comparable to the $Z_c(4430)$, making them all viable molecular candidates.
Given the coexistence of these overlapping resonances, in experimental observations of final states such as $\psi(2S)\pi$ or $D^*\bar D\pi$, significant interference effects---either constructive or destructive---are expected.
However, since the relative phases and production strengths of these poles cannot be determined within the current framework, we do not perform a quantitative analysis of the total line shape that includes interference effects.
Instead, to illustrate the line shape features and open-charm decay ratios characteristic of such broad molecular candidates, we select the first pole in the $D^*\bar D_2^*(^3P_1,^5P_1)$ system ($E=-8.1-76.1i$ MeV) as a representative case.
We denote this pole as \textbf{Pole A} and will discuss its decay process and line shape in the next subsection.

Finally, to validate our single-channel analysis, we performed a full coupled-channel calculation involving all 15 channels (see Table \ref{tab: channels}) for the representative candidate poles identified above.
The resulting energies, component probabilities, and $T$-matrix residues are listed in Table \ref{tab: mutichannel}.
Comparing these results with the single-sector calculations, the pole positions show minimal deviation.
For Pole A (the $Z_c(4430)$ candidate), the properties and $T$-matrix residues are dominated by the $D^*\bar D_2^*(^3P_1,^5P_1)$ channels.
Similarly, for the $\{D\bar D_0^{*}\}$ pole (the $Z_c(4200)$ candidate), the dynamics are dominated by the $\{D\bar D_0^{*}\}(^1P_1)$ channel. In conclusion, our analysis justifies the use of the single-sector approximation---specifically $D^*\bar D_2^*(^3P_1,^5P_1)$ and $\{D\bar D_0^{*}\}(^1P_1)$---for studying these potential exotic states, as the influence of other coupled channels is negligible.

\subsection{Line shape of the $Z_c(4430)$ Candidate}\label{sec:Zcs}

To further investigate the properties of the $Z_c(4430)$ under the molecular hypothesis, we proceed to analyze the decay processes and spectral line shape, assuming it corresponds to the $D^*\bar D_2^*$ resonance (Pole A).
We construct the effective Lagrangian describing the coupling between the resonance ($Z$) and the $D^*\bar D_2^*$/$\bar D^* D_2^*$ channels:
\begin{align}
	\mathcal{L}_Z &= g_{Z_1}\varepsilon^{\alpha\beta\mu\nu}\mathcal{Z}_{\mu\nu}^i\tau^{i}_{ab}\partial_\beta \tilde D_b^{*\gamma\dagger}\tilde D_{2a\alpha\gamma}^{*\dagger} \nonumber\\
	&+ g_{Z_2}\varepsilon^{\alpha\beta\mu\nu}\mathcal{Z}_{\mu\nu}^i\tau^{i}_{ab}\partial^\gamma \tilde D_{b\beta}^{*\dagger}\tilde D_{2a\alpha\gamma}^{*\dagger} + h.c., \label{eq: lageff}
\end{align}
where the field strength tensor is defined as $\mathcal{Z}_{\mu\nu}=\partial_\mu Z_\nu-\partial_\nu Z_\mu$, and $Z_\mu$ represents the field corresponding to the $D^*\bar D_2^*$ molecular state.

Derived from this effective Lagrangian, the energy-dependent partial decay width for the process $Z_c \to D^* \bar{D}_2^*$ is expressed as:
\begin{align}
	\Gamma_{D^* \bar D_2^*}(E) &=\frac{20}{9\pi}m_Z m_{D^*} m_{D_2^*}\bigg[ |g_{Z_1}|^2 + |g_{Z_2}|^2 \frac{m_Z^2}{m_{D_2^*}^2} \nonumber\\
	&\quad - \text{Re}(g_{Z_1}^*g_{Z_2})\frac{m_Z}{m_{D_2^*}} \bigg]p^3 \left(\frac{\Lambda^2}{\Lambda^2+p^2}\right)^2
\end{align}
where $p$ denotes the magnitude of the relative momentum in the center-of-mass frame. To account for the finite size of the hadrons and regularize contributions from the high-momentum region, a monopole form factor $\mathcal{F}(p^2) = \frac{\Lambda^2}{\Lambda^2+p^2}$ is introduced at the interaction vertex.

We adopt the parameters of Pole A listed in Table \ref{tab: results}. The transition amplitude $\langle k_{R,i}|\hat{V}|\phi\rangle$ is extracted via the CSM. By utilizing the ratio of the transition amplitudes into the $(^3P_1)$ and $(^5P_1)$ channels, we determine the coupling ratio $g_{Z_2}/g_{Z_1} = 0.130 - 0.035i$.

Since the traditional Breit-Wigner distribution is often inadequate for such a broad, near-threshold system, we characterize the resonance line shape using a Flatt\'{e}-like parameterization. This approach incorporates an energy-dependent self-energy function $\Sigma(E)$ in the denominator to preserve analyticity. In terms of the bare mass $m_0$ and the coupling constant $g_{Z_1}$, the differential cross-section takes the form:
\begin{equation}
	\frac{d\sigma}{dE} = C \left|\frac{m_0\sqrt{\Gamma_{in}\Gamma_{f}(E)}}{E^2-m_0^2-m_0\Sigma(E)+im_0\Gamma_{D_2^*}(E)}\right|^2, \label{eq:lineshape}
\end{equation}
where $\Gamma_{f}(E)$ is the energy-dependent partial width for a specific final state, and $C$ is a normalization constant. $\Gamma_{in}$ represents the partial width into the incoming channel; since the production mechanism is not specified, we treat it as an effective constant near the $D^*\bar D_2^*$ threshold. $\Gamma_{D_2^*}(E)$ represents the energy-dependent width of the unstable constituent $D_2^*$, as given in Eqs.~\eqref{D2star width1}-\eqref{D2star width2}. Including this term accounts for the additional width contribution inherited from the instability of the $D_2^*$ constituent.

\begin{figure*}[htbp]
	\centering
	\subfigure[]{ \label{fig: lineshape_2_body}
		\includegraphics[width=200pt]{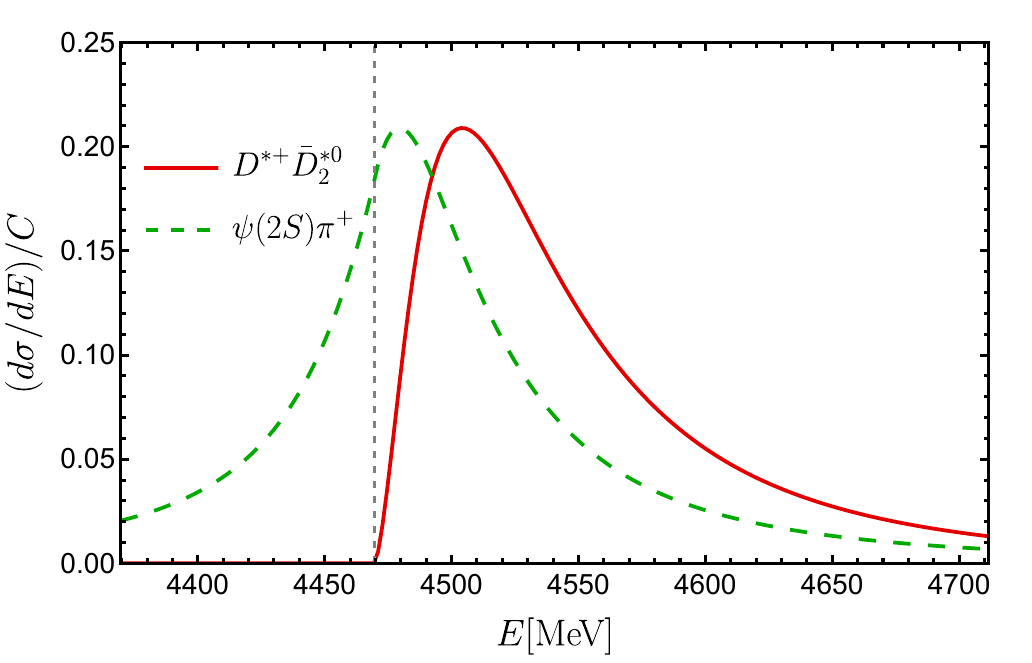}}\hspace{20pt}
	\subfigure[]{ \label{fig: lineshape_3_body}
		\includegraphics[width=200pt]{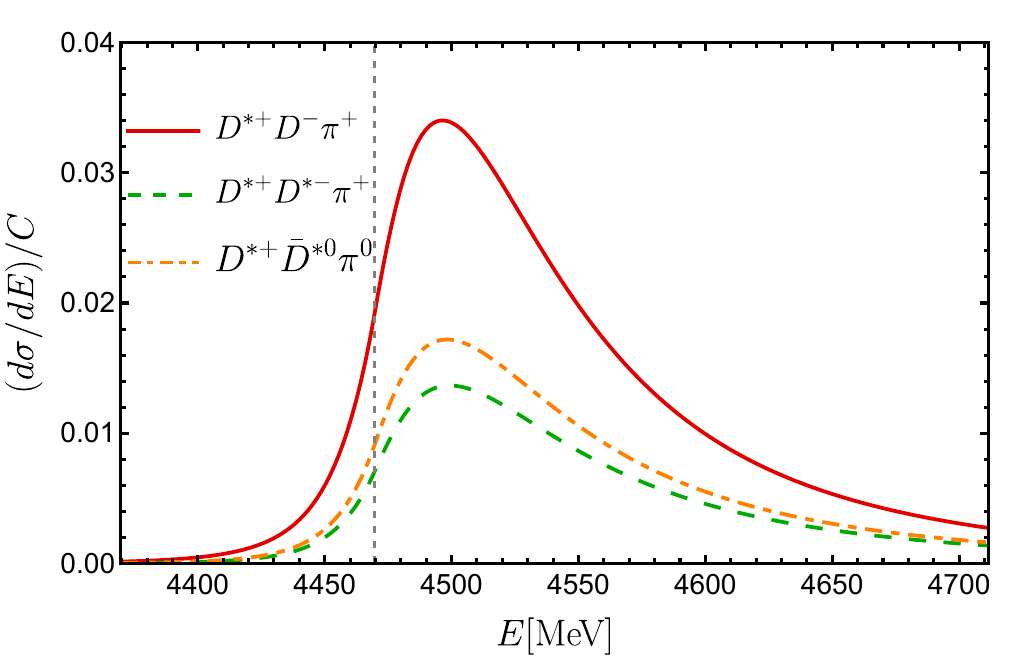}}
	\caption{\small Production line shapes of resonance A (corresponding to the $Z_c(4430)$ candidate) decaying into (a) two-body final states $D^{*+}\bar D_2^{*0}$ and $\psi(2S) \pi^+$, and (b) three-body final states $D^{*+}D^-\pi^+$, $D^{*+}D^{*-}\pi^+$, and $D^{*+}\bar D^{*0}\pi^0$.
		The coupling constant $g_{\psi^\prime}$ is determined by scaling the $\psi(2S) \pi^+$ distribution to share the same peak height as the $D^{*+}\bar D_2^{*0}$ mode, serving solely for shape comparison.}
	\label{fig: lineshape}
\end{figure*}

The imaginary part of the self-energy is determined by the optical theorem:
\begin{align}
	\text{Im} \Sigma(E) &=-\Gamma_{D^* \bar D_2^*}(E).
\end{align}
The real part, $\text{Re} \Sigma(E)$, is then derived from the imaginary part via the unsubtracted dispersion relation:
\begin{align}
	\text{Re} \Sigma(E) &=\frac{1}{\pi}\mathcal{P}\int_{E_{th}}^{\infty}\frac{\text{Im}\Sigma(E^\prime)}{E^\prime-E}dE^\prime,
\end{align}
where $E_{th} = m_{D^*} + m_{D_2^*}$ is the threshold of the open-charm channel. It is worth noting that the form factor introduced in Eq.~\eqref{eq:lineshape} (implicitly in the width) ensures the convergence of the integral at the high-energy limit. This dispersive approach guarantees that the real and imaginary parts of the self-energy satisfy the Kramers-Kronig relations, thereby preserving the analyticity of the scattering amplitude.

We determine the parameters of the Flatt\'{e}-like distribution by aligning the complex pole position with the theoretical energy $E_R$ of Pole A, while simultaneously fitting the peak of the line shape to the experimental central value of the $Z_c(4430)$ \cite{ParticleDataGroup:2024cfk}. This matching procedure leads to the selection of the cutoff parameter $\Lambda=0.35$ GeV and determines the remaining parameters as: $m_0=4526.9$ MeV, $g_{Z_1} = 0.417$ \text{GeV}$^{-5/2}$, and $g_{Z_2} = 0.054 - 0.015i$ \text{GeV}$^{-5/2}$.

Considering the positively charged state $Z_c^+$, the two-body modes are $D^{*+}\bar D_2^{*0}$ and $D_2^{*+}\bar D^{*0}$. The relevant three-body modes include $D^{*+}\bar{D}^{-}\pi^+$, $D^{0}\bar{D}^{*0}\pi^+$, $D^{*+}\bar{D}^0\pi^0$, $D^{+}\bar{D}^{*0}\pi^0$, $D^{*+}\bar{D}^{*-}\pi^+$, $D^{*0}\bar{D}^{*0}\pi^+$, and $D^{*+}\bar{D}^{*0}\pi^0$. Based on isospin symmetry and the decay branching fractions of the excited mesons, we derive the following relations:
\begin{align}
	&\Gamma(A\to D^{*+}\bar D_2^{*0}) = \Gamma(A\to D_2^{*+}\bar D^{*0}), \nonumber\\
	&\Gamma(A\to D^{*+}D^{-}\pi^+) = \Gamma(A\to D^{0}\bar{D}^{*0}\pi^+) \nonumber\\
	&= 2\Gamma(A\to D^{*+}\bar{D}^0\pi^0) = 2\Gamma(A\to D^{+}\bar{D}^{*0}\pi^0), \nonumber\\
	&\Gamma(A\to D^{*+}D^{*-}\pi^+) = \Gamma(A\to D^{*0}\bar{D}^{*0}\pi^+). \label{eq:width 3body}
\end{align}
The $D^{*+}\bar{D}^{*0}\pi^0$ channel involves quantum interference between two sequential decay diagrams:
\begin{align}
	i\mathcal{M} &= i\mathcal{M}(A\to D^{*+}\bar D_{2}^{*0} \to D^{*+}\bar{D}^{*0}\pi^0) \nonumber\\
	&\quad + i\mathcal{M}(A\to D_{2}^{*+}\bar D^{*0} \to D^{*+}\bar{D}^{*0}\pi^0).
\end{align}

In Fig.~\ref{fig: lineshape}, we plot the energy-dependent differential cross-sections, $\frac{1}{C}\frac{d\sigma}{dE}$, for both the two-body and three-body open-charm final states. The incoming width parameter is fixed to the total width of Pole A, i.e., $\Gamma_{in} = 152$ MeV. For comparison, we also include the hidden-charm decay mode $A\to \psi(2S) \pi^+$, assuming an S-wave effective width $\Gamma_{\psi^\prime}(E) = g_{\psi^\prime}|\vec{p}_\pi|$. This channel is of particular interest as it corresponds to the discovery mode of the $Z_c(4430)$ and is theoretically expected to be favored due to the substantial spatial overlap between the extended $P$-wave molecular state and the radially excited $\psi(2S)$ \cite{Liu2014}. Since the effective coupling $g_{\psi^\prime}$ cannot be rigorously determined within the current framework, we calibrate its value by normalizing the peak of the $\psi(2S)\pi^+$ line shape to match the maximum height of the dominant $D^{*+}\bar{D}_2^{*0}$ mode. This approach facilitates a direct comparison of their spectral structures---specifically the width and threshold behaviors---without implying a quantitative prediction of their relative branching fractions. The figure also displays the line shapes for the representative three-body modes $D^{*+}D^-\pi^+$, $D^{*+}D^{*-}\pi^+$, and $D^{*+}\bar D^{*0}\pi^0$. The distributions for other final states can be inferred from the isospin relations in Eq.~\eqref{eq:width 3body}.

The line shapes exhibit distinct features governed by their respective kinematic regimes. For the hidden-charm channel $\psi(2S)\pi^+$, which lies far above its threshold, the phase space factor varies slowly. Consequently, the distribution is dominated by the modulus squared of the resonance propagator. Interestingly, the apparent width of this peak is approximately $80$ MeV, which is noticeably narrower than the intrinsic pole width ($\Gamma \approx 152$ MeV). This narrowing arises from the energy-dependent self-energy $\Sigma(E)$ in the denominator, which is significantly suppressed by the form factor with the soft cutoff ($\Lambda=0.35$ GeV) in the high-energy region.

Similarly, the open-charm line shapes are governed by the interplay between threshold kinematics and the hadronic form factor. The low-energy behavior is suppressed by the $P$-wave phase space ($\propto p^3$), leading to a delayed rise in the cross-section. Meanwhile, the high-energy tail is directly damped by the vertex form factor. This combined effect results in an asymmetric open-charm peak with an apparent width that is comparable to the hidden-charm mode but topologically distinct. It is crucial to emphasize that while the apparent spectral widths in both channels are narrowed by the regularization scheme, the extracted pole width ($\sim 150$ MeV) remains in good agreement with the experimental value ($\sim 180$ MeV).

Finally, the differential cross-sections for the three-body final states (e.g., $D^{*+}\bar{D}^{*0}\pi^0$) shown in Fig.~\ref{fig: lineshape}(b) exhibit similar asymmetric structures to the two-body open-charm mode. These distributions arise from the subsequent decay of the unstable $D_2^*$ component and confirm that the broad resonant signature of the $Z_c(4430)$ candidate is preserved in the experimentally observable multi-body final states.

\section{Summary}\label{sec:summary}

In this study, we employ the OBE potential to investigate the hidden-charm tetraquark states with $I^G(J^{PC})=1^+(1^{+-})$ within the molecular picture. The constituents include the $S$-wave $(D, D^*)$ doublet and the $P$-wave $(D_0^*, D_1^\prime)$ and $(D_1, D_2^*)$ doublets. 
In total, 10 sectors are examined, comprising 7 single-channel systems: $\{D\bar D^*\}(^3S_1)$, $D^*\bar D^*(^3S_1)$, $\{D\bar D_0^{*}\}(^1P_1)$, $\{D\bar D_1^\prime\}(^3P_1)$, $\{D\bar D_1\}(^3P_1)$, $[D\bar D_2^*](^5P_1)$, and $[D^*\bar D_0^*](^3P_1)$; and 3 coupled-channel systems exhibiting spin-mixing: $D^*\bar D_1^\prime(^{1,3,5}P_1)$, $D^*\bar D_1(^{1,3,5}P_1)$, and $D^*\bar D_2^*(^{3,5}P_1)$.

We constructed the effective Lagrangians based on Heavy Quark Symmetry and chiral symmetry. By matching the invariant amplitudes from the quark model with those at the hadronic level, we determined the necessary coupling constants, which show excellent agreement with experimental data. In the study of molecular systems with $Z_c$-like quantum numbers, we observed that the potentials from $\rho$ and $\omega$ exchanges largely cancel each other, leaving the medium- and long-range interactions dominated by $\pi$ and $\sigma$ exchanges.

We solved the Schrödinger equation in momentum space using the CSM and verified the results via the CSLSE. Consequently, we treated the $(D, D^*)$ mesons as stable particles, focusing on the $P$-wave molecular states formed by their combinations with the unstable $(D_0^*, D_1^\prime)$ and $(D_1, D_2^*)$ mesons.
Initially, using $g_\sigma=2.73$, we re-evaluated the $S$-wave molecular candidates corresponding to $T_{cc}$, $X(3872)$, $G(3900)$, and $Z_c(3900)$ from our previous work \cite{Lin2024b}, confirming that the qualitative conclusions remain unchanged.
Subsequently, under the instantaneous approximation ($q_0=0$), we found that a sub-threshold resonance emerges in every $P$-wave system, while systems with spin-mixing exhibit two sub-threshold resonances. These results are consistent with expectations from heavy quark spin symmetry.
Upon incorporating the static limit correction (finite $q_0$), the pole positions shifted, with changes in the imaginary parts on the order of a few MeV.

Finally, the inclusion of the self-energy contributions led to drastic changes in the pole parameters; specifically, the increase in the imaginary parts is positively correlated with the widths of the constituent $P$-wave charmed hadrons.
Notably, the broad resonances identified in the $D^*\bar D_1$, $D^*\bar D_2^*$, $\{D\bar D_0^{*}\}$, and $\{D\bar D_1^\prime\}$ sectors exhibit masses and widths comparable to the exotic states $Z_c(4430)$ and $Z_c(4200)$, suggesting multiple possible molecular assignments.
Taking the $D^*\bar D_2^*(^{3,5}P_1)$ and $\{D\bar D_0^{*}\}(^1P_1)$ sectors as prime examples, we demonstrated that the three-body effects significantly amplify their widths, bringing the calculated parameters into excellent agreement with experimental values.

In Subsection \ref{sec:Zcs}, we analyzed the line shape of the $Z_c(4430)$ candidate (assuming the $D^*\bar D_2^*$ assignment) using a Flatt\'{e}-like parameterization that incorporates self-energy corrections derived from CSM amplitudes.
We compared the line shapes for the open-charm ($D^* \bar D_2^*$ and three-body) and hidden-charm ($\psi(2S)\pi$) channels.
Our results show that the open-charm structures are located above the $D^*\bar D_2^*$ threshold and exhibit distinct asymmetric line shapes governed by the $P$-wave phase space and the form factor.
Crucially, although the apparent spectral widths in both channels are compressed by the regularization scheme, the extracted pole width is consistent with the experimental observation.

In summary, we have applied the OBE model to investigate $P$-wave molecules composed of a negative-parity meson ($D^{(*)}$) and a positive-parity meson ($D_0^*$, $D_1^\prime$, $D_1$, $D_2^*$).
We conclude that these systems universally support broad, near-threshold resonances, and that three-body decay effects play a critical role in determining their widths and pole positions.
The theoretical framework and results presented here, particularly the relations for open-charm decay channels, serve as a valuable guide for future experimental searches for hidden-charm tetraquarks in three-body final states such as $D^*\bar D^*\pi$ and $D^*\bar D\pi$.

	\acknowledgments{This work was supported in part by the National Natural Science Foundation of China under Grants No. 12405160, No. 12547101, No. 11975033, No. 12147168, No. 12070131001, No. 12105072, No.12475137,
	and No.12405088. J. B. Cheng is also supported by
	the Fundamental Research Funds for the Central Universities (Grant
	No. 23CX06061A), the Shandong Provincial Natural Science
	Foundation, China (Grant No. ZR2024QA041). J. Z. Wang is also supported by the Start-up Funds of Chongqing University.
}

\section{Appendix}

\subsection{Appendix A: Expanded Interaction Lagrangians}\label{appendix:lagrangian}

After expanding the Lagrangian in Eq.~\eqref{eq:lagrangian}, we obtain the effective Lagrangians describing the interactions of heavy flavor mesons with light mesons. The effective Lagrangian for pseudoscalar meson ($\pi, \eta$) exchanges is given by:
\begin{align}
	\mathcal{L}_{P^{*}P\mathcal{M}} &= -\frac{2g}{f_\pi} (P_{b}P_{a\mu}^{*\dagger}+P_{b\mu}^{*}P_{a}^{\dagger})\partial^{\mu}\mathcal{M}_{ba}, \nonumber\\
	\mathcal{L}_{P^{*}P^{*}\mathcal{M}} &= i\frac{2g}{f_\pi} \varepsilon_{\alpha\mu\nu\lambda}v^{\alpha}P_{b}^{*\mu}P_{a}^{*\nu\dagger}\partial^{\lambda}\mathcal{M}_{ba}, \nonumber\\
	\mathcal{L}_{P_1^{\prime}P_0^*\mathcal{M}} &= -\frac{2g^\prime}{f_\pi} (P_{1b}^{\prime\mu}P_{0a}^{*\dagger}+P_{0b}^{*}P_{1a}^{\prime\mu\dagger})\partial_{\mu}\mathcal{M}_{ba}, \nonumber\\
	\mathcal{L}_{P_1^{\prime}P_1^{\prime}\mathcal{M}} &= \frac{2ig^{\prime}}{f_\pi} \varepsilon_{\alpha\mu\nu\lambda}v^{\alpha}P_{1b}^{\prime\mu}P_{1a}^{\prime\nu\dagger}\partial^{\lambda}\mathcal{M}_{ba}, \nonumber\\
	\mathcal{L}_{P_1P_1\mathcal{M}} &= -\frac{5ig^{\prime\prime}}{3f_\pi} \varepsilon_{\alpha\mu\nu\lambda}v^{\alpha}P_{1b}^{\mu}P_{1a}^{\nu\dagger}\partial^{\lambda}\mathcal{M}_{ba}, \nonumber\\
	\mathcal{L}_{P_1P_2^*\mathcal{M}} &= -\sqrt{\frac{2}{3}}\frac{g^{\prime\prime}}{f_\pi}(P_{2b}^{*\mu\nu}P_{1a\mu}^{\dagger}+P_{1b}^\mu P_{2a\mu}^{*\nu\dagger})\partial_\nu \mathcal{M}_{ba}, \nonumber\\
	\mathcal{L}_{P_2^{*}P_2^{*}\mathcal{M}} &= \frac{2ig^{\prime\prime}}{f_\pi} \varepsilon_{\alpha\mu\nu\lambda}v^{\alpha}P_{2b}^{*\beta\mu}P_{2a\beta}^{*\nu\dagger}\partial^{\lambda}\mathcal{M}_{ba}, \nonumber\\
	\mathcal{L}_{PP_0^*\mathcal{M}} &= \frac{2f^{\prime\prime}}{f_\pi} (P_{0b}^* P_a^{\dagger}+P_b P_{0a}^{*\dagger})v_\alpha\partial^{\alpha}\mathcal{M}_{ba}, \nonumber\\
	\mathcal{L}_{P^{*}P_1^{\prime}\mathcal{M}} &= -\frac{2f^{\prime\prime}}{f_\pi} (P_{1b}^{\prime\mu} P_{a\mu}^{*\dagger}+P_{b\mu}^{*}P_{1a}^{\prime\mu\dagger})v_\alpha\partial^{\alpha}\mathcal{M}_{ba}, \nonumber\\
	\mathcal{L}_{PP_2^*\mathcal{M}} &= -\frac{2i(h_1+h_2)}{\Lambda_\chi f_\pi} (P_{2b}^{*\mu\nu} P_a^{\dagger}-P_b P_{2a}^{*\mu\nu\dagger})\partial_\mu\partial_\nu\mathcal{M}_{ba}, \nonumber\\
	\mathcal{L}_{P^{*}P_1\mathcal{M}} &= -\sqrt{\frac{2}{3}}\frac{i(h_1+h_2)}{\Lambda_\chi f_\pi}[3g_{\mu\nu}g_{\alpha\beta}+g_{\nu\beta}(-g_{\mu\alpha}+v_\mu v_\alpha)] \nonumber\\
	&\quad \times (P_{1b}^{\nu} P_{a}^{*\beta\dagger}-P_{b}^{*\beta}P_{1a}^{\nu\dagger})\partial^{\mu}\partial^{\alpha}\mathcal{M}_{ba}, \nonumber\\
	\mathcal{L}_{P^{*}P_2^{*}\mathcal{M}} &= \frac{2(h_1+h_2)}{\Lambda_\chi f_\pi} \varepsilon_{\rho\nu\alpha\beta}v^{\rho}\partial_{\mu}\partial^{\alpha}\mathcal{M}_{ba}(P_{2b}^{*\mu\nu}P_{a}^{*\beta\dagger}+P_{b}^{*\beta}P_{2a}^{*\mu\nu\dagger}), \nonumber\\
	\mathcal{L}_{P_0^{*}P_1\mathcal{M}} &= -\sqrt{\frac{2}{3}}\frac{2f^\prime}{f_\pi} (P_{1b}^{\mu}P_{0a}^{*\dagger}+P_{0b}^{*}P_{1a}^{\mu\dagger})\partial_{\mu}\mathcal{M}_{ba}, \nonumber\\
	\mathcal{L}_{P_1^{\prime}P_1\mathcal{M}} &= -\sqrt{\frac{2}{3}}\frac{if^\prime}{f_\pi} \varepsilon_{\alpha\mu\nu\lambda}v^{\alpha}(P_{1b}^{\mu}P_{1a}^{\prime\nu\dagger}+P_{1b}^{\prime\mu}P_{1a}^{\nu\dagger})\partial^{\lambda}\mathcal{M}_{ba}, \nonumber\\
	\mathcal{L}_{P_1^{\prime}P_2\mathcal{M}} &= \frac{2f^\prime}{f_\pi} (P_{2b}^{*\mu\nu}P_{1a\nu}^{\prime\dagger}+P_{1b}^{\prime\nu}P_{2a\nu}^{*\mu\dagger})\partial_{\mu}\mathcal{M}_{ba}.
\end{align}

The effective Lagrangian for scalar meson ($\sigma$) exchanges is given by:
\begin{align}
	\mathcal{L}_{PP\sigma} &= -2g_\sigma P_a^\dagger P_a\sigma, \nonumber\\
	\mathcal{L}_{P^*P^*\sigma} &= 2g_\sigma P_{a\mu}^{*\dagger} P_a^{*\mu} \sigma, \nonumber\\
	\mathcal{L}_{P_0^*P_0^*\sigma} &= 2g^\prime_\sigma P_{0a}^{*\dagger}P_{0a}^*\sigma, \nonumber\\
	\mathcal{L}_{P^\prime_1P^\prime_1\sigma} &= -2g^\prime_\sigma P_{1a\mu}^{\prime\dagger} P^{\prime\mu}_{1a} \sigma, \nonumber\\
	\mathcal{L}_{P_1 P_1\sigma} &= -2g^{\prime\prime}_\sigma P_{1a\mu}^{\dagger} P_{1a}^\mu\sigma, \nonumber\\
	\mathcal{L}_{P_2^* P_2^*\sigma} &= 2g^{\prime\prime}_\sigma P_{2a\mu\nu}^{*\dagger}P_{2a}^{*\mu\nu}\sigma, \nonumber\\
	\mathcal{L}_{PP^\prime_1\sigma} &= -\frac{2ih_\sigma}{f_\pi} (P_a^\dagger P_{1a}^{\prime\mu}-P_{1a}^{\prime\mu\dagger}P_a)\partial_\mu\sigma, \nonumber\\
	\mathcal{L}_{P^* P_0^*\sigma} &= -\frac{i2h_\sigma}{f_\pi} (P_{a}^{*\mu\dagger}P_{0a}^{*}-P_{0a}^{*\dagger}P_{a}^{*\mu})\partial_\mu\sigma, \nonumber\\
	\mathcal{L}_{P^* P^\prime_1\sigma} &= \frac{2h_\sigma}{f_\pi}\varepsilon_{\lambda\mu\nu\alpha}v^\lambda(P_a^{*\mu\dagger}P_{1a}^{\prime\nu}-P_{1a}^{\prime\mu\dagger}P_a^{*\nu})\partial^\alpha\sigma, \nonumber\\
	\mathcal{L}_{PP_1\sigma} &= -2\sqrt{\frac{2}{3}}\frac{ih^\prime_\sigma}{f_\pi} (P_a^\dagger P_{1a}^{\mu}-P_{1a}^{\mu\dagger}P_a)\partial_\mu\sigma, \nonumber\\
	\mathcal{L}_{P^* P_1\sigma} &= -\sqrt{\frac{2}{3}}\frac{h^\prime_\sigma}{f_\pi}\varepsilon_{\lambda\mu\nu\alpha}v^\lambda(P_a^{*\mu\dagger}P_{1a}^\nu-P_{1a}^{\mu\dagger} P_a^{*\nu})\partial^\alpha\sigma, \nonumber\\
	\mathcal{L}_{P^*P_2^*\sigma} &= \frac{2ih^\prime_\sigma}{f_\pi}(P_{a\mu}^{*\dagger}P_{2a}^{*\mu\nu}-P_{2a}^{*\mu\nu\dagger}P_{a\mu}^{*})\partial_\nu\sigma.
\end{align}

The effective Lagrangian for vector meson ($\rho, \omega$) exchanges describing intra-multiplet interactions is:
\begin{align}
	\mathcal{L}_{PP\mathcal{V}} &= -\sqrt{2}\beta g_V P_b P_a^\dagger v\cdot\hat \rho_{ba}, \nonumber\\
	\mathcal{L}_{PP^*\mathcal{V}} &= -2\sqrt{2}\lambda g_V \varepsilon_{\lambda\alpha\mu\nu}v^\lambda (P_b P_a^{*\alpha\dagger}+P_b^{*\alpha}P_a^\dagger) \partial^\mu \hat\rho_{ba}^\nu, \nonumber\\
	\mathcal{L}_{P^*P^*\mathcal{V}} &= \sqrt2 \beta g_V P_b^{*\mu}P_{a\mu}^{*\dagger} v\cdot\hat\rho_{ba} \nonumber\\
	&\quad -2\sqrt2 i\lambda g_V P_b^{*\mu}P_a^{*\nu\dagger}(\partial_\mu \hat\rho_\nu-\partial_\nu \hat\rho_\mu)_{ba}, \nonumber\\
	\mathcal{L}_{P_0^* P_0^*\mathcal{V}} &= \sqrt{2}\beta_1 g_V P_{0b}^* P_{0a}^{*\dagger} v\cdot\hat \rho_{ba}, \nonumber\\
	\mathcal{L}_{P_0^* P^\prime_1\mathcal{V}} &= 2\sqrt{2}\lambda_1 g_V \varepsilon_{\lambda\alpha\mu\nu}v^\lambda (P_{0b}^* P_{1a}^{\prime\alpha\dagger}+P_{1b}^{\prime\alpha}P_{0a}^{*\dagger})  \partial^\mu \hat\rho_{ba}^\nu, \nonumber\\
	\mathcal{L}_{P^\prime_1 P^\prime_1\mathcal{V}} &= -\sqrt2 \beta_1 g_V P_{1b}^{\prime\mu}P_{1a\mu}^{\prime\dagger} v\cdot\hat\rho_{ba} \nonumber\\
	&\quad +2\sqrt2 i\lambda_1 g_V P_{1b}^{\prime\mu}P_{1a}^{\prime\nu\dagger}(\partial_\mu \hat\rho_\nu-\partial_\nu \hat\rho_\mu)_{ba}, \nonumber\\
	\mathcal{L}_{P_1 P_1 \mathcal{V}} &= -\sqrt2\beta_2 g_V P_{1b}^\mu P_{1a\mu}^\dagger (v\cdot\hat\rho)_{ba} \nonumber\\
	&\quad +i\frac{5}{3}\sqrt{2}\lambda_2 g_V P_{1b}^\mu P_{1a}^{\nu\dagger} (\partial_\mu\hat \rho_\nu-\partial_\nu\hat \rho_\mu)_{ba}, \nonumber\\
	\mathcal{L}_{P_1 P_2^*\mathcal{V}} &= -\frac{2}{\sqrt{3}}\lambda_2 g_V \varepsilon_{\lambda\alpha\mu\nu}v^\lambda \nonumber\\
	&\quad \times \Big[ (P_{2b}^{*\sigma\alpha} P_{1a\sigma}^{\dagger}+P_{1b\sigma}P_{2a}^{*\sigma\alpha\dagger}) (\partial^\mu \hat \rho^\nu)_{ba} \nonumber\\
	&\quad -(P_{2b}^{*\sigma\mu} P_{1a}^{\nu\dagger}+P_{1b}^{\nu}P_{2a}^{*\sigma\mu\dagger})(\partial_\sigma \hat \rho^\alpha)_{ba} \Big], \nonumber\\
	\mathcal{L}_{P_2^* P_2^*\mathcal{V}} &= \sqrt2 \beta_2 g_V P_{2b}^{*\mu\nu}P_{2a\mu\nu}^{*\dagger} (v\cdot\hat\rho)_{ba} \nonumber\\
	&\quad -2\sqrt2 i\lambda_2 g_V P_{2b}^{*\sigma\mu}P_{2a\sigma}^{*\nu\dagger}(\partial_\mu \hat\rho_\nu-\partial_\nu \hat\rho_\mu)_{ba}.
\end{align}

The mixing terms between the ground-state doublets and the $P$-wave excited doublets are:
\begin{align}
	\mathcal{L}_{PP^\prime_1\mathcal{V}} &= -\sqrt{2}\zeta g_V (P_b P_{1a}^{\prime\mu\dagger}+P_{1b}^{\prime\mu}P_a^\dagger) (\hat\rho_{\mu})_{ba} \nonumber\\
	&\quad -2\sqrt{2}i\mu g_V v^\mu (P_b P_{1a}^{\prime\nu\dagger}-P_{1b}^{\prime\nu} P_a^\dagger) \nonumber\\
	&\quad \times (\partial_\mu \hat\rho_\nu-\partial_\nu \hat\rho_\mu)_{ba}, \nonumber\\
	\mathcal{L}_{P^*P_0^*\mathcal{V}} &= -\sqrt2 \zeta g_V (P_b^{*\mu}P_{0a}^{*\dagger}+P_{0b}^{*}P_a^{*\mu\dagger}) (\hat\rho_\mu)_{ba} \nonumber\\
	&\quad -2\sqrt2 i\mu g_V v^\mu(P_b^{*\nu}P_{0a}^{*\dagger}-P_{0b}^{*}P_a^{*\nu\dagger}) \nonumber\\
	&\quad \times (\partial_\mu \hat\rho_\nu-\partial_\nu \hat\rho_\mu)_{ba}, \nonumber\\
	\mathcal{L}_{P^*P^\prime_1\mathcal{V}} &= \sqrt2 \zeta g_V i\varepsilon_{\lambda\mu\nu\alpha}v^\lambda(P_b^{*\mu}P_{1a}^{\prime\nu\dagger}+P_{1b}^{\prime\mu}P_a^{*\nu\dagger}) (\hat\rho^\alpha)_{ba} \nonumber\\
	&\quad -\sqrt2 \mu g_V \varepsilon_{\alpha\beta\mu\nu}(P_b^{*\alpha}P_{1a}^{\prime\beta\dagger}-P_{1b}^{\prime\alpha}P_a^{*\beta\dagger}) \nonumber\\
	&\quad \times (\partial^\mu \hat\rho^\nu-\partial^\nu \hat\rho^\mu)_{ba}, \nonumber\\
	\mathcal{L}_{PP_1\mathcal{V}} &= -\frac{2}{\sqrt{3}}\zeta_1 g_V (P_{1b}^\mu P_a^\dagger+P_b P_{1a}^{\mu\dagger})(\hat\rho_\mu)_{ba} \nonumber\\
	&\quad -i\frac{2}{\sqrt{3}}\mu_1 g_V v^\mu(P_{1b}^\nu P_a^\dagger-P_b P_{1a}^{\nu\dagger}) \nonumber\\
	&\quad \times (\partial_\mu \hat\rho_\nu-\partial_\nu \hat\rho_\mu)_{ba}, \nonumber\\
	\mathcal{L}_{P^*P_1\mathcal{V}} &= -\frac{1}{\sqrt{3}} \zeta_1 g_V i\varepsilon_{\lambda\mu\nu\alpha}v^\lambda (P_{1b}^{\mu}P_{a}^{*\nu\dagger}+P_{b}^{*\mu}P_{1a}^{\nu\dagger}) (\hat\rho^\alpha)_{ba} \nonumber\\
	&\quad +\frac{1}{\sqrt3}\mu_1 g_V v_\mu v^\sigma \varepsilon_{\sigma\nu\alpha\beta}(P_{1b}^{\alpha}P_{a}^{*\beta\dagger}-P_{b}^{*\alpha}P_{1a}^{\beta\dagger}) \nonumber\\
	&\quad \times (\partial^\mu \hat\rho^\nu-\partial^\nu \hat\rho^\mu)_{ba}, \nonumber\\
	\mathcal{L}_{P^*P_2^*\mathcal{V}} &= \sqrt2 \zeta_1 g_V (P_{2b}^{*\mu\nu}P_{a\nu}^{*\dagger}+P_{b\nu}^{*}P_{2a}^{*\mu\nu\dagger})(\hat\rho_\mu)_{ba} \nonumber\\
	&\quad -i\sqrt2 \mu_1 g_V v^\mu(P_{2b}^{*\nu\alpha}P_{a\alpha}^{*\dagger}-P_{b\alpha}^{*}P_{2a}^{*\nu\alpha\dagger}) \nonumber\\
	&\quad \times (\partial_\mu \hat\rho_\nu-\partial_\nu \hat\rho_\mu)_{ba}.
\end{align}

Finally, the mixing terms between different $P$-wave multiplets are:
\begin{align}
	\mathcal{L}_{P_0^*P_1\mathcal{V}} &= \frac{2}{\sqrt{3}}\mu_2 g_V \varepsilon_{\alpha\beta\mu\nu}v^{\alpha}(P_{1b}^\beta P_{0a}^{*\dagger}+P_{0b}^* P_{1a}^{\beta\dagger})\partial^\mu \hat\rho^\nu, \nonumber\\
	\mathcal{L}_{P^\prime_1 P_1\mathcal{V}} &= -i\frac{1}{\sqrt3}\mu_2 g_V (P_{1b}^\mu P_{1a}^{\prime\nu\dagger}+P_{1b}^{\prime\mu}P_{1a}^{\nu\dagger}) \nonumber\\
	&\quad \times (\partial_\mu \hat\rho_\nu-\partial_\nu \hat\rho_\mu)_{ba}, \nonumber\\
	\mathcal{L}_{P^\prime_1 P_2^*\mathcal{V}} &= -\sqrt2 \mu_2 g_V \varepsilon_{\sigma\nu\alpha\beta}v^\sigma(P_{2b\mu}^{*\alpha}P_{1a}^{\prime\beta\dagger}+P_{1b}^{\prime\beta}P_{2a\mu}^{\prime\alpha\dagger}) \nonumber\\
	&\quad \times (\partial^\mu \hat\rho^\nu-\partial^\nu \hat\rho^\mu)_{ba}.
\end{align}

The interaction terms not explicitly listed are zero.

The Lagrangians for the anti-charmed hadrons can be derived from those of the charmed sector via the charge conjugation ($C$) transformation. The transformation properties for the heavy meson fields are given by:
\begin{equation}
	\begin{aligned}
		P_a &\xrightarrow{C} \tilde{P}_a, & P_a^* &\xrightarrow{C} -\tilde{P}_a^*, \\
		P_{0a}^* &\xrightarrow{C} \tilde{P}_{0a}^*, & P_{1a}^\prime &\xrightarrow{C} \tilde{P}_{1a}^\prime, \\
		P_{1a} &\xrightarrow{C} \tilde{P}_{1a}, & P_{2a}^* &\xrightarrow{C} -\tilde{P}_{2a}^*.
	\end{aligned}
\end{equation}
For the light mesons, the fields transform as follows:
\begin{equation}
	\mathcal{M} \xrightarrow{C} \mathcal{M}^T, \quad \hat{\rho} \xrightarrow{C} -\hat{\rho}^T, \quad \sigma \xrightarrow{C} \sigma.
\end{equation}

\subsection{Appendix B: coupling constants}\label{sec: couplings}

In this appendix, we detail the determination of the coupling constants associated with the $(0^-,1^-)$, $(0^+,1^+)$, and $(1^+,2^+)$ heavy meson doublets. We employ the quark model combined with experimental data to systematically estimate these parameters.

The effective Lagrangian describing the interactions of the light constituent quarks ($u, d$) is given by
\begin{eqnarray}
	&&\mathcal{L}_{q}=-\frac{g_A}{\sqrt 2 f_\pi}\bar \psi \gamma^\mu \gamma_5 \partial_\mu(\pi^i\tau^i)\psi-g_{\sigma}^q \bar\psi \sigma\psi \nonumber\\
	&&-g_{\rho}^q \bar\psi \gamma^\mu(\rho_\mu^i \tau^i+\omega_\mu)\psi-f_\rho^q \bar \psi \sigma^{\mu\nu}\partial_\mu(\rho_\nu^i \tau^i+\omega_\nu)\psi.
\end{eqnarray}
Under the assumption that the heavy quark acts as a spectator and does not participate in the coupling with light mesons, we can derive the relations between the coupling constants at the hadronic and quark levels by matching the transition matrix elements calculated in both frameworks. Taking the classic process $D^*\to D\pi$ as an example, the invariant amplitude at the hadronic level is
\begin{eqnarray}
	&&i\mathcal{M}_H[D^{*0}(m_J=0)\to D^{+}(m'_J=0)\pi^{-}]\nonumber\\
	&&=\frac{g}{f_\pi}\epsilon_\mu(m_J=0) q^{\mu}=-\frac{g}{f_\pi}q^3, \label{eq:iMh}
\end{eqnarray}
where $m_J$ ($m'_J$) denotes the third component of the total angular momentum of the initial (final) heavy meson, and $\vec{q}$ represents the pion momentum. On the other hand, the invariant amplitude at the quark level is expressed as
\begin{eqnarray}
	&&i\mathcal{M}_Q(\bar u\to\bar d \pi^-)=\frac{g_A}{f_\pi}\left[\chi_{\bar u}^\dagger\left(\vec\sigma\cdot\vec q\right)\chi_{\bar d}\right]\nonumber\\
	&&-\frac{g_A}{f_\pi}q^0\left[\chi_{\bar u}^\dagger \vec\sigma\cdot\left(\frac{\vec{p}^{\prime}}{2m_d}+\frac{\vec p}{2m_u}\right)\chi_{\bar d}\right],\label{eq:iMq}
\end{eqnarray}
where $\chi_{\bar u}$ ($\chi_{\bar d}$) is the Pauli spinor of the $\bar u$ ($\bar d$) antiquark, and $m_u$ ($m_d$) and $\vec{p}$ ($\vec{p}^{\prime}$) are the mass and momentum of the $\bar u$ ($\bar d$) antiquark in the heavy quark rest frame, satisfying $\vec{q}=\vec{p}-\vec{p}^{\prime}$. Since $|q^0| \ll m_{u} \approx 400$ MeV for the $D^*\to D\pi$ process, the first term dominates the amplitude in Eq.~\eqref{eq:iMq}. Consequently, the corresponding amplitude at the quark level becomes
\begin{eqnarray}
	&&i\mathcal{M}_Q[D^{*0}(m_J=0)\to D^{+}(m'_J=0)\pi^{-}]\nonumber\\
	&&=\langle\Psi_{m'_J=0}(D^+)|\frac{g_A}{f_\pi}(\vec\sigma\cdot\vec q)|\Psi_{m_J=0}(D^{*0})\rangle,\label{eq:iMq2}
\end{eqnarray}
where the wave function of the $D^*$ meson is decomposed as $|\Psi_{m_J}(D^{*0})\rangle=\sum_{lm\lambda}|\psi_{lm}(D^{*0})\rangle |\chi_{\lambda}(D^{*0})\rangle$. Specifically, the spatial part is $|\psi_{lm}(D^{*0})\rangle=|\psi_{00}(D^{*0})\rangle$, and the spin part is $|\chi_{\lambda}(D^{*0})\rangle =|\chi_{0}(D^{*0})\rangle = \frac{1}{\sqrt{2}}(\uparrow\downarrow+\downarrow\uparrow)\otimes c\bar u$. The operator $\vec{\sigma}\cdot\vec{q}$ acts only on the spin wave function, leading to
\begin{align}
	\langle\vec\sigma\cdot\vec q\rangle &= \sum_{lm\lambda;l'm'\lambda'}\langle\psi_{l'm'}(D^{+})|\psi_{lm}(D^{*0})\rangle\nonumber\\
	&\langle\chi_{\lambda'}(D^{+})|\vec\sigma\cdot\vec q|\chi_{\lambda}(D^{*0})\rangle \nonumber\\
	&=\int \frac{d^3p}{(2\pi)^3}\psi_{00}^*(\vec{p}',D^{+})\psi_{00}(\vec{p},D^{*0}) (q^3). \label{eq:iMq3}
\end{align}
Considering the limited phase space of the $D^*\to D\pi$ process, we have $|\vec{q}| \ll |\vec{p}|, |\vec{p}'|$, where $|\vec{p}| \approx 400$ MeV in the heavy quark effective theory \cite{Manohar2000}. This implies $\vec{p}' \approx \vec{p}$ and $\psi_{00}^*(\vec{p}',D^{+}) \approx \psi_{00}^*(\vec{p},D^{+})$. In the heavy quark limit, we assume identical spatial wave functions for the $D$ and $D^*$ mesons, yielding the overlap integral $\int \frac{d^3p}{(2\pi)^3}\psi_{00}^*(\vec{p},D^{+})\psi_{00}(\vec{p},D^{*0})=1$. By matching the hadronic level amplitude in Eq.~\eqref{eq:iMh} with the quark level result derived from Eqs.~\eqref{eq:iMq2} and \eqref{eq:iMq3}, we obtain
\begin{equation}
	g=-g_A,
\end{equation}
where $g_A=0.75$ is taken from the chiral quark model \cite{Manohar:1983md}.

We proceed to determine the remaining coupling constants in the OPE interaction.
For transitions between two heavy mesons with the same parity, the second term in Eq.~\eqref{eq:iMq} is proportional to $\frac{q^0}{2m_u}\langle \vec\sigma\cdot\vec q\rangle$, which is suppressed relative to the first term $\langle \vec\sigma\cdot\vec q\rangle$ due to the factor $q^0/2m_u$. Consequently, we can estimate the coupling constants $g^\prime$, $g^{\prime\prime}$, and $f^\prime$ using a similar approach. We again approximate $\vec{p}^\prime \approx \vec{p}$ in the final state spatial wave function $\psi_{l^\prime m^\prime}^*(\vec{p}^\prime)$ to simplify the integral.
In Ref. \cite{Song2015}, the authors employed a modified Godfrey-Isgur (GI) model where charmed mesons are described by single spherical harmonic oscillator functions $\Psi_{nlm}^{SHO}(\vec p)$. The quantum numbers $n ^{2S+1}L_J$ for the $(D,D^*)$, $(D_0^*,D_1^\prime)$, and $(D_1,D_2^*)$ doublets are assigned as $(1^1S_0, 1^3S_1)$, $(1^3P_0, 1^3P_1)$, and $(1^1P_1, 1^3P_2)$, respectively. Neglecting the mass differences (or scale parameter variations \cite{Song2015}) within the doublets, the wave functions of heavy mesons with the same parity satisfy the orthonormality condition:
\begin{equation}
	\langle\psi_{l^\prime m^\prime}(H^\prime_f)|\psi_{lm}(H_i)\rangle \approx \delta_{l^\prime,l}\delta_{m^\prime,m}.
\end{equation}
Based on this, we derive the following relations:
\begin{equation}
	g^\prime=g_A/3, \quad g^{\prime\prime}=g_A, \quad f^\prime=-\frac{2}{\sqrt3}g_A.
\end{equation}

For transitions involving heavy mesons with different parity, the first term $\langle \vec\sigma\cdot\vec q\rangle$ in Eq.~\eqref{eq:iMq} vanishes due to the orthogonality of the spherical harmonic functions, provided $|\vec{q}| \ll |\vec{p}|, |\vec{p}^\prime|$. However, this condition is not strictly satisfied here. For instance, in the $D_0^*\to D\pi$ decay, $|\vec{q}|\approx 400$ MeV, which is comparable to the estimated internal momentum $|\vec p|$. Thus, this term could yield a non-negligible contribution. Nevertheless, for the purpose of a rapid estimation in this work, we retain the approximation $\vec{p}^\prime \approx \vec{p}$. Under this assumption, the amplitude in Eq.~\eqref{eq:iMq} is dominated by the $\vec p$-dependent term:
\begin{equation}
	i\mathcal{M}_Q(\bar u\to\bar d \pi^-)=-\frac{g_A}{f_\pi}\frac{q^0}{m_u}\left[\chi_{\bar u}^\dagger\left(\vec\sigma\cdot\vec{p}\right)\chi_{\bar d}\right].\label{eq:iMq_diff}
\end{equation}
To simplify the subsequent expressions for coupling constants, we define a general overlap integral $\mathcal{I}(H_a, H_b)$ as:
\begin{equation}
	\mathcal{I}(H_a, H_b) = \frac{1}{(2\pi)^3} \int_0^\infty dp \, p^3 \phi_{L_a}(p, H_a) \phi_{L_b}^*(p, H_b), \label{eq:integral_def}
\end{equation}
where $\phi_L(p,H)$ denotes the radial wave function of hadron $H$ in momentum space. By calculating the spin operator matrix element of $\langle \vec{\sigma}\rangle$ and performing the solid angle integration over $\vec{p}$, we obtain:
\begin{align}
	f^{\prime\prime} &= \frac{1}{3}\frac{g_A}{m_u} \mathcal{I}(D_0^*, D), \nonumber\\
	\frac{h_1+h_2}{\Lambda_\chi} &= \frac{1}{4\sqrt3} \frac{g_A}{m_u^2} \mathcal{I}(D_2^*, D).
\end{align}
Using the constituent quark mass $m_q=340$ MeV and the model parameters corresponding to $\mu=0.03$ from Table II of Ref. \cite{Song2015}, we calculate:
\begin{equation}
	f^{\prime\prime}=1.00 g_A, \qquad (h_1+h_2)/\Lambda_\chi=0.89 g_A \text{ GeV}^{-1}.\label{eq:fpphh}
\end{equation}

Alternatively, $f^{\prime\prime}$ and $(h_1+h_2)/\Lambda_\chi$ can be extracted from the experimental widths of $D_{0}^*$ and $D_1$. Assuming the open-charm channels dominate their widths, i.e., $\Gamma_{D_0^*}=\frac{3}{2}\Gamma(D_0^*\to D\pi^+)$ and $\Gamma_{D_1}=\frac{3}{2}\Gamma(D_1\to D^*\pi^+)$, and using the average masses from Table \ref{tab: mass meson}, we obtain $|f^{\prime\prime}|=0.53$ and $|(h_1+h_2)/\Lambda_\chi|=1.22$ GeV$^{-1}$. Additionally, averaging the values extracted from $D_1$ and $D_2^*$ yields $|(h_1+h_2)/\Lambda_\chi|=1.03$ GeV$^{-1}$. Evidently, these phenomenological values are consistent with the quark model predictions.

The final coupling constants for pion exchange are determined by combining the magnitude from experimental data with the sign predicted by the quark model. Consequently, the adopted values are $g=-0.59$ (magnitude from \cite{ahmedFirstMeasurement2001}), $g^\prime=0.25$, $g^{\prime\prime}=0.75$, $f^\prime=-0.87$, $f^{\prime\prime}=0.53$, and $(h_1+h_2)/\Lambda_\chi=1.03$ GeV$^{-1}$.

For vector meson interactions, the couplings are derived similarly using the definition in Eq.~\eqref{eq:integral_def}:
\begin{eqnarray}
	&&\lambda g_V=-f_\rho^q, \quad \lambda_1 g_V=-f_\rho^q/3,\quad \lambda_2 g_V=f_\rho^q, \nonumber\\
	&&\beta g_V=-2 g_\rho^q, \quad \beta_1 g_V=2 g_\rho^q, \quad \beta_2 g_V=2 g_\rho^q, \nonumber\\
	&&\zeta g_V= -\frac{2}{3}\frac{g_\rho^q}{m_u}\mathcal{I}(D_1^\prime, D), \quad \mu g_V= \frac{2}{3}\frac{f_\rho^q}{m_u}\mathcal{I}(D_1^\prime, D), \nonumber\\
	&&\zeta_1 g_V= -\frac{2}{\sqrt3}\frac{g_\rho^q}{m_u}\mathcal{I}(D_1, D), \quad \mu_1 g_V= \frac{2}{\sqrt3}\frac{f_\rho^q}{m_u}\mathcal{I}(D_1, D),\nonumber\\
	&&\mu_2 g_V=4 f_\rho^q/\sqrt{3}.
\end{eqnarray}
We take $g_V=5.8$, $\beta=0.9$, and $\lambda=0.56$ GeV$^{-1}$ \cite{bandoNonlinearRealizationHidden1988,isolaCharmingPenguinContributions2003} as inputs to determine the other vector coupling constants: $\lambda_1=0.19$ GeV$^{-1}$, $\lambda_2=-0.56$ GeV$^{-1}$, $\beta_1=-0.90$, $\beta_2=-0.90$, $\mu=-1.00$ GeV$^{-1}$, $\mu_1=-1.74$ GeV$^{-1}$, $\mu_2=-1.29$ GeV$^{-1}$, $\zeta=0.81$ and $\zeta_1=1.40$.

For scalar meson interactions, we obtain:
\begin{align}
	&g_\sigma=g_\sigma^q, \quad g^\prime_\sigma=-g_\sigma^q, \quad g^{\prime\prime}_\sigma=-g_\sigma^q, \nonumber\\
	&h_\sigma=\frac{1}{6}\frac{f_\pi g_\sigma^q}{m_u^2}\mathcal{I}(D_1^\prime, D), \quad h^\prime_\sigma=-\frac{1}{4\sqrt3}\frac{f_\pi g_\sigma^q}{m_u^2}\mathcal{I}(D_1, D).
\end{align}
Adopting the chiral quark model \cite{Manohar:1983md} and the Goldberger-Treiman relation \cite{Vogl:1991qt}, we set $g_\sigma^q=g_{\pi qq}\approx m_q g_A/f_\pi=2.73$. The corresponding coupling constants are $g_\sigma=2.73$, $g^\prime_\sigma=-2.73$, $g^{\prime\prime}_\sigma=-2.73$, $h_\sigma=0.48$, and $h^\prime_\sigma=-0.48$.

\subsection{Appendix C: potential}\label{appendix:potential}

For simplicity, we first derive the OBE potential for a system of two charmed mesons (e.g., $DD^*$) prior to partial wave decomposition. Subsequently, we apply the G-parity transformation to both the heavy and light mesons to obtain the complete potential for the hidden-charm sector. Taking the $PP^*\to PP^*$ process as an illustrative example, the OBE potential is given by:
\begin{align}
	& V(DD^*\to DD^*) = V_\pi^D+V_\pi^C+V_\eta+V_\eta^D+V_\eta^C+V_\sigma^D \nonumber\\
	& \qquad +V_\sigma^C+V_\rho^D+V_\rho^C+V_\omega^D+V_\omega^C,\nonumber\\
	& V_\pi^D=V_\eta^D=V_\sigma^C=0,\nonumber\\
	& V_\pi^C=\frac{g^2}{2f_\pi^2}\frac{(\vec q \cdot \vec \epsilon_3^\dagger)(\vec q \cdot \vec \epsilon_1)}{q^2-m_\pi^2}\vec \tau_1 \cdot \vec \tau_2,\nonumber\\
	& V_\eta^C=\frac{g^2}{6f_\pi^2}\frac{(\vec q \cdot \vec \epsilon_3^\dagger)(\vec q \cdot \vec \epsilon_1)}{q^2-m_\eta^2}, \quad V_\sigma^D=g_\sigma^2 \frac{\vec\epsilon_4^\dagger\cdot \vec\epsilon_2}{q^2-m_\sigma^2},\nonumber\\
	& V_\rho^D=-\frac{1}{4}\beta^2 g_V^2 \frac{\vec\epsilon_4^\dagger\cdot \vec\epsilon_2}{q^2-m_\rho^2}\vec{\tau}_1\cdot\vec{\tau}_2,\nonumber\\
	& V_\rho^C=\lambda^2 g_V^2 \frac{\vec q^2(\vec\epsilon_3^\dagger\cdot \vec\epsilon_2)-(\vec q\cdot \vec\epsilon_3^\dagger)(\vec q\cdot \vec\epsilon_2)}{q^2-m_\rho^2}\vec{\tau}_1\cdot\vec{\tau}_2,\nonumber\\
	& V_\omega^D=-\frac{1}{4}\beta^2 g_V^2 \frac{\vec\epsilon_4^\dagger\cdot \vec\epsilon_2}{q^2-m_\omega^2},\nonumber\\
	& V_\omega^C=\lambda^2 g_V^2 \frac{\vec q^2(\vec\epsilon_3^\dagger\cdot \vec\epsilon_2)-(\vec q\cdot \vec\epsilon_3^\dagger)(\vec q\cdot \vec\epsilon_2)}{q^2-m_\omega^2}.
\end{align}
where $\vec\epsilon$ ($\vec\epsilon^\dagger$) denotes the polarization vector of the initial (final) $D^*$ meson. $V^D$ and $V^C$ correspond to the direct diagram ($DD^*\to DD^*$) and the cross diagram ($DD^*\to D^*D$), respectively.

We then consider the G-parity transformations: $D\xrightarrow{G}-\bar D$, $D^*\xrightarrow{G}\bar D^*$, $D_0^*\xrightarrow{G}-\bar D_0^*$, $D_1^\prime\xrightarrow{G}-\bar{D}_1^\prime$, $D_1\xrightarrow{G}-\bar{D}_1$, $D_2^*\xrightarrow{G}\bar{D}_2^*$, $\pi\xrightarrow{G}-\pi$, $\eta\xrightarrow{G}\eta$, $\sigma\xrightarrow{G}\sigma$, $\rho\xrightarrow{G}\rho$, and $\omega\xrightarrow{G}-\omega$.
Taking the first channel $\{D\bar D^*\}(^3S_1)$ in Table \ref{tab: channels} as an example, the flavor configuration $\{A\bar B\}=(A\bar B+B\bar A)/\sqrt{2}$ or $[A\bar B]=(A\bar B-B\bar A)/\sqrt{2}$ introduces a phase factor $\delta_C$ in the cross-channel diagrams under G-parity transformation. This factor depends on the symmetry of the final state: $\delta_C=1$ for $\{A\bar B\}$ and $\delta_C=-1$ for $[A\bar B]$. For the process $\{D\bar D^*\}(^3S_1)\to\{D\bar D^*\}(^3S_1)$, we have $\delta_C=1$. Consequently, the OBE potential components transform as:
\begin{widetext}
	% 开启跨页断行功能，确保公式可以自然流转到下一页
	\allowdisplaybreaks
	\begin{align}
		V(D\bar D^*\to D\bar D^*) &= \delta_C\frac{g^2}{2f_\pi^2}(\vec q \cdot \vec \epsilon_3^\dagger)(\vec q \cdot \vec \epsilon_2)\left(\frac{\vec\tau_1\cdot\vec \tau_2}{q^2-m_\pi^2}-\frac{1/3}{q^2-m_\eta^2}\right)+g_\sigma^2 \frac{\vec\epsilon_4^\dagger\cdot \vec\epsilon_2}{q^2-m_\sigma^2}\nonumber\\
		&\quad +g_V^2 \left\{-\frac{1}{4}\beta^2 (\vec\epsilon_4^\dagger\cdot \vec\epsilon_2)-\delta_C\lambda^2 \left[\vec q^2(\vec\epsilon_3^\dagger\cdot \vec\epsilon_2)-(\vec q\cdot \vec\epsilon_3^\dagger)(\vec q\cdot \vec\epsilon_2)\right]\right\} \left(\frac{\vec{\tau}_1\cdot\vec{\tau}_2}{q^2-m_\rho^2}-\frac{1}{q^2-m_\omega^2}\right),\label{eq:V11}\\[10pt]
		V(D^*\bar D^*\to D^*\bar D^*) &= -\frac{g^2}{2f_\pi^2}[\vec q\cdot(i\vec\epsilon_3^\dagger\times\vec\epsilon_1)][\vec q\cdot(i\vec\epsilon_4^\dagger\times\vec\epsilon_2)]\left(\frac{\vec\tau_1\cdot\vec \tau_2}{q^2-m_\pi^2}-\frac{1/3}{q^2-m_\eta^2}\right)+g_\sigma^2\frac{(\vec\epsilon_3^\dagger\cdot\vec\epsilon_1)(\vec\epsilon_4^\dagger\cdot\vec\epsilon_2)}{q^2-m_\sigma^2}\nonumber\\
		&\quad +g_V^2\left\{-\frac{1}{4}\beta^2 (\vec\epsilon_3^\dagger\cdot\vec\epsilon_1)(\vec\epsilon_4^\dagger\cdot\vec\epsilon_2)+\lambda^2 \{\vec {q}^2(i\vec\epsilon_3^\dagger\times\vec\epsilon_1)\cdot(i\vec\epsilon_4^\dagger\times\vec\epsilon_2)-[\vec q\cdot(i\vec\epsilon_3^\dagger\times\vec\epsilon_1)][\vec q\cdot(i\vec\epsilon_4^\dagger\times\vec\epsilon_2)]\}\right\}\nonumber\\
		&\quad \times \left(\frac{\vec{\tau}_1\cdot\vec{\tau}_2}{q^2-m_\rho^2}-\frac{1}{q^2-m_\omega^2}\right),\label{eq:V22}\\[10pt]
		V(D\bar D_0^*\to D\bar D_0^*) &= -\delta_C\frac{f''^2}{2f_\pi^2}q_{0C}^2\left(\frac{\vec\tau_1\cdot\vec \tau_2}{q^2-m_\pi^2}-\frac{1/3}{q^2-m_\eta^2}\right)-g_\sigma g'_\sigma\frac{1}{q^2-m_\sigma^2}+\frac{1}{4}\beta\beta_1 g_V^2\left(\frac{\vec{\tau}_1\cdot\vec{\tau}_2}{q^2-m_\rho^2}-\frac{1}{q^2-m_\omega^2}\right),\label{eq:V33}\\[10pt]
		V(D\bar D'_1\to D\bar D'_1) &= -g_\sigma g'_\sigma\frac{\vec\epsilon_4^\dagger\cdot\vec\epsilon_2}{q^2-m_\sigma^2}+\delta_C\frac{h_\sigma^2}{f_\pi^2}\frac{(\vec q\cdot\vec\epsilon_3^\dagger)(\vec q\cdot\vec\epsilon_2)}{q^2-m_\sigma^2}+\frac{1}{4}\beta\beta_1 g_V^2(\vec\epsilon_4^\dagger\cdot\vec\epsilon_2)\left(\frac{\vec{\tau}_1\cdot\vec{\tau}_2}{q^2-m_\rho^2}-\frac{1}{q^2-m_\omega^2}\right)\nonumber\\
		&\quad +\delta_C g_V^2\left\{\frac{1}{4}(\zeta-2\mu q_{0C})^2 (\vec\epsilon_3^\dagger\cdot\vec\epsilon_2) -\mu^2 (\vec q\cdot\vec\epsilon_3^\dagger)(\vec q\cdot\vec\epsilon_2)\right\}\left(\frac{\vec{\tau}_1\cdot\vec{\tau}_2}{q^2-m_\rho^2}-\frac{1}{q^2-m_\omega^2}\right),\label{eq:V44}\\[10pt]
		V(D\bar D_1\to D\bar D_1) &= -g_\sigma g''_\sigma\frac{\vec\epsilon_4^\dagger\cdot\vec\epsilon_2}{q^2-m_\sigma^2}+\delta_C\frac{2}{3}\frac{h_\sigma^{'2}}{f_\pi^2}\frac{(\vec q\cdot\vec\epsilon_3^\dagger)(\vec q\cdot\vec\epsilon_2)}{q^2-m_\sigma^2}-\frac{1}{4}\beta\beta_2 g_V^2(\vec\epsilon_4^\dagger\cdot\vec\epsilon_2)\left(\frac{\vec{\tau}_1\cdot\vec{\tau}_2}{q^2-m_\rho^2}-\frac{1}{q^2-m_\omega^2}\right)\nonumber\\
		&\quad +\delta_C\frac{g_V^2}{6}\left\{(\zeta_1+\mu q_{0C})^2 (\vec\epsilon_3^\dagger\cdot\vec\epsilon_2) -\mu_1^2(\vec q\cdot\vec\epsilon_3^\dagger)(\vec q\cdot\vec\epsilon_2)\right\}\left(\frac{\vec{\tau}_1\cdot\vec{\tau}_2}{q^2-m_\rho^2}-\frac{1}{q^2-m_\omega^2}\right),\label{eq:V55}\\[10pt]
		V(D\bar D_2^*\to D\bar D_2^*) &= \delta_C\frac{(h_1+h_2)^2}{2\Lambda_\chi^2 f_\pi^2}(\eta_{3ij}^{\dagger}q_i q_j)(\eta_{2lm}^{\dagger}q_l q_m)\left(\frac{\vec\tau_1\cdot\vec\tau_2}{q^2-m_\pi^2}-\frac{1/3}{q^2-m_\eta^2}\right)-g_\sigma g''_\sigma \frac{\eta_{2ij}\eta_{4ij}^\dagger}{q^2-m_\sigma^2}\nonumber\\
		&\quad +\frac{1}{4}\beta\beta_2 g_V^2(\eta_{2ij}\eta_{4ij}^\dagger)\left(\frac{\vec{\tau}_1\cdot\vec{\tau}_2}{q^2-m_\rho^2}-\frac{1}{q^2-m_\omega^2}\right),\label{eq:V66}\\[10pt]
		V(D^*\bar D_0^*\to D^*\bar D_0^*) &= -g_\sigma g'_\sigma\frac{\vec\epsilon_3^\dagger\cdot\vec\epsilon_1}{q^2-m_\sigma^2}-\delta_C\frac{h_\sigma^2}{f_\pi^2}\frac{(\vec q\cdot\vec\epsilon_4^\dagger)(\vec q\cdot\vec\epsilon_1)}{q^2-m_\sigma^2}+\frac{1}{4}\beta\beta_1 g_V^2(\vec\epsilon_3^\dagger\cdot\vec\epsilon_1)\left(\frac{\vec{\tau}_1\cdot\vec{\tau}_2}{q^2-m_\rho^2}-\frac{1}{q^2-m_\omega^2}\right)\nonumber\\
		&\quad -\delta_C g_V^2\left\{\frac{1}{4}(\zeta-2\mu q_{0C})^2 (\vec\epsilon_4^\dagger\cdot\vec\epsilon_1) -\mu^2 (\vec q\cdot\vec\epsilon_4^\dagger)(\vec q\cdot\vec\epsilon_1)\right\}\left(\frac{\vec{\tau}_1\cdot\vec{\tau}_2}{q^2-m_\rho^2}-\frac{1}{q^2-m_\omega^2}\right),\label{eq:V77}\\[10pt]
		V(D^*\bar D'_1\to D^*\bar D'_1) &= -\frac{gg'}{2f_\pi^2}[\vec q\cdot(i\vec\epsilon_3^\dagger\times\vec\epsilon_1)][\vec q\cdot(i\vec\epsilon_4^\dagger\times\vec\epsilon_2)]\left(\frac{\vec\tau_1\cdot\vec\tau_2}{q^2-m_\pi^2}-\frac{1/3}{q^2-m_\eta^2}\right)\nonumber\\
		&\quad +\delta_C\frac{f''^2}{2f_\pi^2}q_{0C}^{2}(\vec\epsilon_3^\dagger\cdot\vec\epsilon_1)(\vec\epsilon_4^\dagger\cdot\vec\epsilon_2)\left(\frac{\vec\tau_1\cdot\vec\tau_2}{q^2-m_\pi^2}-\frac{1/3}{q^2-m_\eta^2}\right)-g_\sigma g'_\sigma\frac{(\vec\epsilon_3^\dagger\cdot\vec\epsilon_1)(\vec\epsilon_4^\dagger\cdot\vec\epsilon_2)}{q^2-m_\sigma^2}\nonumber\\
		&\quad -\delta_C \frac{h_\sigma^2}{f_\pi^2}\frac{[\vec q\cdot(i\vec\epsilon_3^\dagger\times\vec\epsilon_1)][\vec q\cdot(i\vec\epsilon_4^\dagger\times\vec\epsilon_2)]}{q^2-m_\sigma^2} +g_V^2\Bigg\{\frac{1}{4}\beta\beta_1(\vec\epsilon_3^\dagger\cdot\vec\epsilon_1)(\vec\epsilon_4^\dagger\cdot\vec\epsilon_2)\nonumber\\
		&\quad -\lambda\lambda_1\{\vec {q}^2(i\vec\epsilon_3^\dagger\times\vec\epsilon_1)\cdot(i\vec\epsilon_4^\dagger\times\vec\epsilon_2)-[\vec q\cdot(i\vec\epsilon_3^\dagger\times\vec\epsilon_1)][\vec q\cdot(i\vec\epsilon_4^\dagger\times\vec\epsilon_2)]\}\Bigg\}\left(\frac{\vec{\tau}_1\cdot\vec{\tau}_2}{q^2-m_\rho^2}-\frac{1}{q^2-m_\omega^2}\right)\nonumber\\
		&\quad -\delta_C g_V^2\left\{\frac{1}{4}(\zeta-2\mu q_{0C})^2 [(i\vec\epsilon_3^\dagger\times\vec\epsilon_1)\cdot(i\vec\epsilon_4^\dagger\times\vec\epsilon_2)]-\mu^2 [\vec q\cdot(i\vec\epsilon_3^\dagger\times\vec\epsilon_1)][\vec q\cdot(i\vec\epsilon_4^\dagger\times\vec\epsilon_2)]\right\}\nonumber\\
		&\quad \times \left(\frac{\vec{\tau}_1\cdot\vec{\tau}_2}{q^2-m_\rho^2}-\frac{1}{q^2-m_\omega^2}\right),\label{eq:V88}\\[10pt]
		V(D^*\bar D_1\to D^*\bar D_1) &= \delta_C\frac{3}{2}\frac{(h_1+h_2)^2}{2\Lambda_\chi^2 f_\pi^2}\left[(\epsilon_{3i}^\dagger\epsilon_{1j}-\frac{1}{3}\delta_{ij}\vec{\epsilon}_3^\dagger\cdot\vec{\epsilon}_1)q_i q_j\right] \left[(\epsilon_{4l}^\dagger\epsilon_{2m}-\frac{1}{3}\delta_{lm}\vec{\epsilon}_4^\dagger\cdot\vec{\epsilon}_2)q_l q_m\right]\left(\frac{\vec\tau_1\cdot\vec\tau_2}{q^2-m_\pi^2}-\frac{1/3}{q^2-m_\eta^2}\right)\nonumber\\
		&\quad -\frac{5}{6}\frac{gg''}{2f_\pi^2}[\vec q\cdot(i\vec\epsilon_3^\dagger\times\vec\epsilon_1)][\vec q\cdot(i\vec\epsilon_4^\dagger\times\vec\epsilon_2)]\left(\frac{\vec\tau_1\cdot\vec\tau_2}{q^2-m_\pi^2}-\frac{1/3}{q^2-m_\eta^2}\right)\nonumber\\
		&\quad -g_\sigma g''_\sigma\frac{(\vec\epsilon_3^\dagger\cdot\vec\epsilon_1)(\vec\epsilon_4^\dagger\cdot\vec\epsilon_2)}{q^2-m_\sigma^2}-\delta_C \frac{h_\sigma^{'2}}{6f_\pi^2}\frac{[\vec q\cdot(i\vec\epsilon_3^\dagger\times\vec\epsilon_1)][\vec q\cdot(i\vec\epsilon_4^\dagger\times\vec\epsilon_2)]}{q^2-m_\sigma^2}\nonumber\\
		&\quad +g_V^2\left\{\frac{1}{4}\beta\beta_2(\vec\epsilon_3^\dagger\cdot\vec\epsilon_1)(\vec\epsilon_4^\dagger\cdot\vec\epsilon_2)+\frac{1}{2}\lambda\lambda_2\{\vec {q}^2(i\vec\epsilon_3^\dagger\times\vec\epsilon_1)\cdot(i\vec\epsilon_4^\dagger\times\vec\epsilon_2)-[\vec q\cdot(i\vec\epsilon_3^\dagger\times\vec\epsilon_1)][\vec q\cdot(i\vec\epsilon_4^\dagger\times\vec\epsilon_2)]\}\right\}\nonumber\\
		&\quad \times \left(\frac{\vec{\tau}_1\cdot\vec{\tau}_2}{q^2-m_\rho^2}-\frac{1}{q^2-m_\omega^2}\right)-\frac{1}{24}\delta_C g_V^2\Big\{(\zeta_1+\mu_1 q_{0C})^2 [(i\vec\epsilon_3^\dagger\times\vec\epsilon_1)\cdot(i\vec\epsilon_4^\dagger\times\vec\epsilon_2)]\nonumber\\
		&\quad -\mu_1^2 [\vec q\cdot(i\vec\epsilon_3^\dagger\times\vec\epsilon_1)][\vec q\cdot(i\vec\epsilon_4^\dagger\times\vec\epsilon_2)]\Big\}\left(\frac{\vec{\tau}_1\cdot\vec{\tau}_2}{q^2-m_\rho^2}-\frac{1}{q^2-m_\omega^2}\right),\label{eq:V99}\\[10pt]
		V(D^*\bar D_2^*\to D^*\bar D_2^*) &= \frac{gg''}{2f_\pi^2}\left\{[\vec{q}\cdot(i\vec{\epsilon}_3^\dagger\times\vec\epsilon_1)](i\varepsilon_{ijk}\eta_{4li}^\dagger\eta_{2lj}q_k)\right\}\left(\frac{\vec\tau_1\cdot\vec\tau_2}{q^2-m_\pi^2}-\frac{1/3}{q^2-m_\eta^2}\right)-g_\sigma g''_\sigma \frac{(\vec\epsilon_3^\dagger\cdot\vec\epsilon_1)(\eta_{4ij}^\dagger\eta_{2ij})}{q^2-m_\sigma^2}\nonumber\\
		&\quad +\delta_C \frac{(h_1+h_2)^2}{2\Lambda_\chi^2 f_\pi^2}(i\varepsilon_{ijk}\epsilon_{1k}\eta_{3in}^\dagger q_j q_n)(i\varepsilon_{i'j'k'}\epsilon_{4k'}^\dagger\eta_{2in'}q_{j'}q_{n'})\left(\frac{\vec\tau_1\cdot\vec\tau_2}{q^2-m_\pi^2}-\frac{1/3}{q^2-m_\eta^2}\right)\nonumber\\
		&\quad +\delta_C \frac{h_\sigma^{'2}}{f_\pi^2}\frac{(q_i\eta_{3ij}^\dagger\epsilon_{1j})(q_l\eta_{2lm}\epsilon_{4m}^\dagger)}{q^2-m_\sigma^2}+g_V^2\Bigg\{\frac{1}{4}\beta\beta_2(\vec\epsilon_3^\dagger\cdot\vec\epsilon_1)(\eta_{4ij}^\dagger\eta_{2ij})\nonumber\\
		&\quad +\lambda\lambda_2 [\vec{q}\times(\vec\epsilon_3^\dagger\times\vec\epsilon_1)]_k (\eta_{4ik}^\dagger\eta_{2ij}-\eta_{4ij}^\dagger\eta_{2ik})q_j\Bigg\}\left(\frac{\vec{\tau}_1\cdot\vec{\tau}_2}{q^2-m_\rho^2}-\frac{1}{q^2-m_\omega^2}\right)\nonumber\\
		&\quad +\delta_C\frac{g_V^2}{4}\left\{(\zeta_1-\mu_1 q_{0C})^2(\eta_{3ik}^\dagger\epsilon_{1i})(\eta_{2jk}\epsilon_{4j}^\dagger)-\mu_1^2(q_i\eta_{3ij}^\dagger\epsilon_{1j})(q_l\eta_{2lm}\epsilon_{4m}^\dagger)\right\}\nonumber\\
		&\quad \times \left(\frac{\vec{\tau}_1\cdot\vec{\tau}_2}{q^2-m_\rho^2}-\frac{1}{q^2-m_\omega^2}\right).\label{eq:V1010}
	\end{align}
\end{widetext}

In our formalism, $\bm{\tau}$ denotes the isospin operator represented by the Pauli matrices. We adopt the phase convention where the isospin doublets correspond to the quark pair $(u, d)$ and the antiquark pair $(\bar{d}, -\bar{u})$. Accordingly, the ground state charmed and anti-charmed meson doublets are defined as $(D^{(*)+}, -D^{(*)0})^\mathrm{T}$ and $(\bar{D}^{(*)0}, D^{(*)-})^\mathrm{T}$, respectively. This convention is consistently applied to the excited $P$-wave doublets. Under this definition, the isospin factor is given by $\langle \bm{\tau}_1 \cdot \bm{\tau}_2 \rangle = 2I(I+1)-3$. For the isovector $Z_c$ states ($I=1$) investigated in this work, this yields $\langle \bm{\tau}_1 \cdot \bm{\tau}_2 \rangle = 1$.

Here, $\epsilon^\mu$ represents the polarization vector for $D^*$, $D_1^\prime$, and $D_1$ mesons, defined as:
\begin{eqnarray}
	&&\epsilon^{\mu}(\pm1)=\mp\frac{1}{\sqrt2}(0,1,\pm i,0)^\mu,\quad \epsilon^{\mu}(0)=(0,0,0,1)^\mu.\nonumber
\end{eqnarray}
The polarization tensor $\eta^{\mu\nu}$ for the $D_2^*$ meson is given by:

\begin{align}
	\eta^{\mu\nu}(\lambda=\pm2) &= \frac{1}{2}
	\begin{pmatrix}
		0 & 0 & 0 & 0 \\
		0 & 1 & \pm i & 0 \\
		0 & \pm i & -1 & 0 \\
		0 & 0 & 0 & 0
	\end{pmatrix}^{\!\mu\nu}, \nonumber\\
	\eta^{\mu\nu}(\lambda=\pm1) &= \mp\frac{1}{2}
	\begin{pmatrix}
		0 & 0 & 0 & 0 \\
		0 & 0 & 0 & 1 \\
		0 & 0 & 0 & \pm i \\
		0 & 1 & \pm i & 0
	\end{pmatrix}^{\!\mu\nu}, \nonumber\\
	\eta^{\mu\nu}(\lambda=0) &= \frac{1}{\sqrt 6}
	\begin{pmatrix}
		0 & 0 & 0 & 0 \\
		0 & -1 & 0 & 0 \\
		0 & 0 & -1 & 0 \\
		0 & 0 & 0 & 2
	\end{pmatrix}^{\!\mu\nu}.
\end{align}

\begin{widetext}
	\allowdisplaybreaks
	The diagonal potentials for channels 1 to 8 after partial wave decomposition are:
	\begin{align}
		V_{11}(p',p) &= 2\pi\int_{-1}^{1}dt\Bigg\{\frac{g^2}{2f_\pi^2}\frac{1}{3}(p'^2+p^2-2p'pt)G(\pi,\eta,q_{0C},p',p,t)+g_\sigma^2 G(m_\sigma,q_{0D},p',p,t) \nonumber\\
		&\quad -\frac{1}{4}g_V^2\beta^2 G(\rho,\omega,q_{0D},p',p,t) -g_V^2\lambda^2\frac{2}{3}(p'^2+p^2-2p'pt) G(\rho,\omega,q_{0C},p',p,t)\Bigg\},\label{eq:V11_diag}\\[10pt]
		V_{22}(p',p) &= 2\pi\int_{-1}^{1}dt\Bigg\{\frac{g^2}{2f_\pi^2}\frac{1}{3}(p'^2+p^2-2p'pt)G(\pi,\eta,q_{0C},p',p,t)+g_\sigma^2 G(m_\sigma,q_{0D},p',p,t)\nonumber\\
		&\quad -\frac{1}{4}g_V^2 \beta^2 G(\rho,\omega,q_{0D},p',p,t)-g_V^2\lambda^2 \frac{2}{3}(p'^2+p^2-2p'pt) G(\rho,\omega,q_{0C},p',p,t)\Bigg\},\label{eq:V22_diag}\\[10pt]
		V_{33}(p',p) &= 2\pi\int_{-1}^{1}tdt\Bigg\{-\frac{f''^2}{2f_\pi^2}q_{0C}^2 G(\pi,\eta,q_{0C},p',p,t)-g_\sigma g'_\sigma G(m_\sigma,q_{0D},p',p,t) +\frac{1}{4}g_V^2\beta\beta_1 G(\rho,\omega,q_{0D},p',p,t)\Bigg\},\label{eq:V33_diag}\\[10pt]
		V_{44}(p',p) &= 2\pi\int_{-1}^{1}dt\Bigg\{-g_\sigma g'_\sigma t G(m_\sigma,q_{0D},p',p,t)+\frac{h_\sigma^2}{f_\pi^2}\frac{1}{2}p'p(1-t^2)G(m_\sigma,q_{0C},p',p,t)\nonumber\\
		&\quad +\frac{1}{4}g_V^2\beta\beta_1 t G(\rho,\omega,q_{0D},p',p,t)+g_V^2\left[\frac{1}{4}(\zeta-2\mu q_{0C})^2 t-\mu^2\frac{1}{2}p'p(1-t^2)\right]  G(\rho,\omega,q_{0C},p',p,t)\Bigg\},\label{eq:V44_diag}\\[10pt]
		V_{55}(p',p) &= 2\pi\int_{-1}^{1}dt\Bigg\{-g_\sigma g''_\sigma t G(m_\sigma,q_{0D},p',p,t)+\frac{2}{3}\frac{h_\sigma^{'2}}{f_\pi^2}\frac{1}{2}p'p(1-t^2)G(m_\sigma,q_{0C},p',p,t)\nonumber\\
		&\quad -\frac{1}{4}g_V^2\beta\beta_2 t G(\rho,\omega,q_{0D},p',p,t)+\frac{g_V^2}{6}\left[(\zeta_1+\mu q_{0C})^2 t-\mu_1^2\frac{1}{2}p'p(1-t^2)\right]  G(\rho,\omega,q_{0C},p',p,t)\Bigg\},\label{eq:V55_diag}\\[10pt]
		V_{66}(p',p) &= 2\pi\int_{-1}^{1}dt\Bigg\{
		-\frac{(h_1+h_2)^2}{2\Lambda_\chi^2 f_\pi^2}\frac{1}{15}\Big[4(p'^4+p^4)t+2p'^2 p^2 t(7+5t^2)-(p'^3 p+p' p^3)(3+13t^2)\Big]\nonumber\\
		&\quad \times G(\pi,\eta,q_{0C},p',p,t) -g_\sigma g''_\sigma t G(m_\sigma,q_{0D},p',p,t) +\frac{1}{4}\beta\beta_2 g_V^2 t G(\rho,\omega,q_{0D},p',p,t)\Bigg\},\label{eq:V66_diag}\\[10pt]
		V_{77}(p',p) &= 2\pi\int_{-1}^{1}dt\Bigg\{-g_\sigma g'_\sigma t G(m_\sigma,q_{0D},p',p,t)+\frac{h_\sigma^2}{f_\pi^2}\frac{1}{2}p'p(1-t^2) G(m_\sigma,q_{0C},p',p,t) \nonumber\\
		&\quad +\frac{1}{4}\beta\beta_1 g_V^2 t G(\rho,\omega,q_{0D},p',p,t)+g_V^2\left[\frac{1}{4}(\zeta-2\mu q_{0C})^2 t-\mu^2\frac{1}{2}p'p(1-t^2) \right] G(\rho,\omega,q_{0C},p',p,t)\Bigg\},\label{eq:V77_diag}
	\end{align}
	The propagator functions are defined as:
	\begin{align}
		G(m,q_0,p^\prime,p,t) &= \frac{-1}{p^{\prime 2}+p^2-2p^\prime pt+m^2-q_0^2}, \nonumber\\
		G(\pi,\eta,q_0,p^\prime,p,t) &= G(m_\pi,q_0,p^\prime,p,t) -\frac{1}{3}G(m_\eta,q_0,p^\prime,p,t), \nonumber\\
		G(\rho,\omega,q_0,p^\prime,p,t) &= G(m_\rho,q_0,p^\prime,p,t) -G(m_\omega,q_0,p^\prime,p,t).
	\end{align}
\end{widetext}

\begin{widetext}
	\allowdisplaybreaks
		The partial wave projected potentials for $D^*\bar{D}_1^\prime\to D^*\bar{D}_1^\prime$ (channels 8--10) are:
		\begin{align}
			V_{88}(p',p) &= 2\pi\int_{-1}^{1}tdt\Bigg\{-\frac{gg'}{2f_\pi^2}\frac{-2}{3}(p'^2+p^2-2p'pt)G(\pi,\eta,q_{0D},p',p,t)-\frac{f''^2}{2f_\pi^2}q_{0C}^{2} G(\pi,\eta,q_{0C},p',p,t)\nonumber\\
			&\quad -g_\sigma g'_\sigma G(m_\sigma,q_{0D},p',p,t)+ \frac{h_\sigma^2}{f_\pi^2}\frac{-2}{3}(p'^2+p^2-2p'pt)G(m_\sigma,q_{0C},p',p,t)\nonumber\\
			&\quad +g_V^2\left[\frac{1}{4}\beta\beta_1+\lambda\lambda_1\frac{4}{3}(p'^2+p^2-2p'pt)\right] G(\rho,\omega,q_{0D},p',p,t)\nonumber\\
			&\quad + g_V^2 \left[-\frac{1}{2}(\zeta-2\mu q_{0C})^2 +\mu^2\frac{2}{3}(p'^2+p^2-2p'pt)\right]G(\rho,\omega,q_{0C},p',p,t)\Bigg\},\label{eq:V88}\\[5pt]
			V_{99}(p',p) &= 2\pi\int_{-1}^{1}dt\Bigg\{\frac{gg'}{2f_\pi^2}\frac{1}{2}(1-t^2)G(\pi,\eta,q_{0D},p',p,t)+\frac{f''^2}{2f_\pi^2}q_{0C}^{2} t G(\pi,\eta,q_{0C},p',p,t)\nonumber\\
			&\quad -g_\sigma g'_\sigma t G(m_\sigma,q_{0D},p',p,t)+ \frac{h_\sigma^2}{f_\pi^2}\frac{1}{2}(1-t^2)G(m_\sigma,q_{0C},p',p,t)\nonumber\\
			&\quad +g_V^2\left[\frac{1}{4}\beta\beta_1 t-\lambda\lambda_1\left(-(p'^2+p^2)t+p'p\frac{1}{2}(1+3t^2)\right)\right]G(\rho,\omega,q_{0D},p',p,t)\nonumber\\
			&\quad -g_V^2 \left[-\frac{1}{4}(\zeta-2\mu q_{0C})^2 t +\mu^2\frac{1}{2}(1-t^2)\right]G(\rho,\omega,q_{0C},p',p,t)\Bigg\},\label{eq:V99}\\[5pt]
			V_{10,10}(p',p) &= 2\pi\int_{-1}^{1}dt\Bigg\{\frac{gg'}{2f_\pi^2}\frac{1}{30}\left[4(p'^2+p^2)t+p'p(-21+13t^2)\right]G(\pi,\eta,q_{0D},p',p,t)\nonumber\\
			&\quad -\frac{f''^2}{2f_\pi^2}q_{0C}^{2} t G(\pi,\eta,q_{0C},p',p,t)-g_\sigma g'_\sigma t G(m_\sigma,q_{0D},p',p,t)\nonumber\\
			&\quad -\frac{h_\sigma^2}{f_\pi^2}\frac{1}{30}\left[4(p'^2+p^2)t+p'p(-21+13t^2)\right]G(m_\sigma,q_{0C},p',p,t)\nonumber\\
			&\quad +g_V^2\left[\frac{1}{4}\beta\beta_1 t-\lambda\lambda_1\frac{1}{30}\left(34(p'^2+p^2)t-p'p(21+47t^2)\right)\right]G(\rho,\omega,q_{0D},p',p,t)\nonumber\\
			&\quad +g_V^2 \left[\frac{1}{4}(\zeta-2\mu q_{0C})^2 t+\mu^2\frac{1}{30}\left(4(p'^2+p^2)t+p'p(-21+13t^2)\right)\right] G(\rho,\omega,q_{0C},p',p,t)\Bigg\},\label{eq:V1010_diag}\\[5pt]
			V_{89}(p',p) &= 0,\quad V_{9,10}(p',p)=0,\nonumber\\[5pt]
			V_{8,10}(p',p) &= 2\pi\int_{-1}^{1}dt\Bigg\{-\frac{gg'}{2f_\pi^2}\frac{1}{3\sqrt5}\left[2(p'^2+p^2)t-p'p(3+t^2)\right]G(\pi,\eta,q_{0D},p',p,t)\nonumber\\
			&\quad +\frac{h_\sigma^2}{f_\pi^2}\frac{1}{3\sqrt5}\left[2(p'^2+p^2)t-p'p(3+t^2)\right]G(m_\sigma,q_{0C},p',p,t)\nonumber\\
			&\quad +g_V^2\lambda\lambda_1 \frac{1}{3\sqrt5}\left[2(p'^2+p^2)t-p'p(3+t^2)\right] G(\rho,\omega,q_{0D},p',p,t)\nonumber\\
			&\quad -g_V^2 \mu^2 \frac{1}{3\sqrt5}\left[2(p'^2+p^2)t-p'p(3+t^2)\right] G(\rho,\omega,q_{0C},p',p,t)\Bigg\}.\label{eq:V810}
		\end{align}
\end{widetext}

\begin{widetext}
	\allowdisplaybreaks
		The partial wave projected potentials for $D^*\bar{D}_1\to D^*\bar{D}_1$ (channels 11--13) are:
		\begin{align}
			V_{11,11}(p',p) &= 2\pi\int_{-1}^{1}tdt\Bigg\{-\frac{1}{3}\frac{(h_1+h_2)^2}{2\Lambda_\chi^2 f_\pi^2}(p'^2+p^2-2p'pt)^2 G(\pi,\eta,q_{0C},p',p,t)+\frac{5}{6}\frac{gg''}{2f_\pi^2}\frac{2}{3}(p'^2+p^2-2p'pt)\nonumber\\
			&\quad \times G(\pi,\eta,q_{0D},p',p,t)-g_\sigma g''_\sigma G(m_\sigma,q_{0D},p',p,t)- \frac{h_\sigma^{'2}}{6f_\pi^2}\frac{2}{3}(p'^2+p^2-2p'pt)G(m_\sigma,q_{0C},p',p,t)\nonumber\\
			&\quad +g_V^2\left[\frac{1}{4}\beta\beta_2-\frac{1}{2}\lambda\lambda_2\frac{4}{3}(p'^2+p^2-2p'pt)\right]G(\rho,\omega,q_{0D},p',p,t)\nonumber\\
			&\quad +\frac{1}{24} g_V^2\left[-2(\zeta_1+\mu_1 q_{0C})^2+\mu_1^2\frac{2}{3}(p'^2+p^2-2p'pt)\right]G(\rho,\omega,q_{0C},p',p,t)\Bigg\},\label{eq:V1111}\\[5pt]
			V_{12,12}(p',p) &= 2\pi\int_{-1}^{1}dt\Bigg\{\frac{3}{2}\frac{(h_1+h_2)^2}{2\Lambda_\chi^2 f_\pi^2}\frac{1}{18}\Big[-4(p'^4+p^4)t-2p'^2 p^2 t(7+5t^2)\nonumber\\
			&\quad +(p'^3 p+p'p^3)(3+13t^2)\Big] G(\pi,\eta,q_{0C},p',p,t)+\frac{5}{6}\frac{gg''}{2f_\pi^2}\frac{1}{2}p'p(1-t^2) G(\pi,\eta,q_{0D},p',p,t)\nonumber\\
			&\quad -g_\sigma g''_\sigma t G(m_\sigma,q_{0D},p',p,t)+ \frac{h_\sigma^{'2}}{6f_\pi^2}\frac{1}{2}p'p(1-t^2)G(m_\sigma,q_{0C},p',p,t)\nonumber\\
			&\quad +g_V^2\left[\frac{1}{4}\beta\beta_2 t+\frac{1}{2}\lambda\lambda_2\left(-(p'^2+p^2)t+p'p\frac{1}{2}(1+3t^2)\right)\right]G(\rho,\omega,q_{0D},p',p,t)\nonumber\\
			&\quad -\frac{1}{24} g_V^2\left[-(\zeta_1+\mu_1 q_{0C})^2 t+\mu_1^2\frac{1}{2}p'p(1-t^2)\right]G(\rho,\omega,q_{0C},p',p,t)\Bigg\},\label{eq:V1212}\\[5pt]
			V_{13,13}(p',p) &= 2\pi\int_{-1}^{1}dt\Bigg\{-\frac{3}{2}\frac{(h_1+h_2)^2}{2\Lambda_\chi^2 f_\pi^2}\frac{1}{30}p'p(p'^2+p^2-2p'pt)(1-t^2) G(\pi,\eta,q_{0C},p',p,t) \nonumber\\
			&\quad +\frac{5}{6}\frac{gg''}{2f_\pi^2}\frac{1}{30}\left[4(p'^2+p^2)t+p'p(-21+13t^2)\right] G(\pi,\eta,q_{0D},p',p,t) \nonumber\\
			&\quad-g_\sigma g''_\sigma t G(m_\sigma,q_{0D},p',p,t)-\frac{h_\sigma^{'2}}{6f_\pi^2}\frac{1}{30}\left[4(p'^2+p^2)t +p'p(-21+13t^2)\right] G(m_\sigma,q_{0C},p',p,t)\nonumber\\
			&\quad +g_V^2\Bigg[\frac{1}{4}\beta\beta_2 t+\frac{1}{2}\lambda\lambda_2\frac{1}{30}\Big(34(p'^2+p^2)t
			 -p'p(21+47t^2)\Big)\Bigg]G(\rho,\omega,q_{0D},p',p,t)\nonumber\\
			&\quad +\frac{1}{24} g_V^2\Bigg[(\zeta_1+\mu_1 q_{0C})^2 t+\mu_1^2\frac{1}{30}\left(4(p'^2+p^2)t+p'p(-21+13t^2)\right)\Bigg]G(\rho,\omega,q_{0C},p',p,t)\Bigg\},\label{eq:V1313}\\[5pt]
			V_{11,12}(p',p) &= 0,\quad V_{12,13}(p',p)=0,\nonumber\\[5pt]
			V_{11,13}(p',p) &= 2\pi\int_{-1}^{1}dt\Bigg\{-\frac{3}{2}\frac{(h_1+h_2)^2}{2\Lambda_\chi^2 f_\pi^2}\frac{1}{9\sqrt5}\Big[2(p'^4+p^4)t+2p'^2 p^2 t(5+t^2)-(p'^3 p+p'p^3)(3+5t^2)\Big] G(\pi,\eta,q_{0C},p',p,t)\nonumber\\
			&\quad -\frac{5}{6}\frac{gg''}{2f_\pi^2}\frac{1}{3\sqrt5}\left[2(p'^2+p^2)t-p'p(3+t^2)\right] G(\pi,\eta,q_{0D},p',p,t)+\frac{h_\sigma^{'2}}{6f_\pi^2}\frac{1}{3\sqrt5}\left[2(p'^2+p^2)t-p'p(3+t^2)\right]\nonumber\\
			&\quad \times G(m_\sigma,q_{0C},p',p,t) -\frac{1}{2}g_V^2\lambda\lambda_2\frac{1}{3\sqrt5}\left[2(p'^2+p^2)t-p'p(3+t^2)\right]G(\rho,\omega,q_{0D},p',p,t)\nonumber\\
			&\quad -\frac{1}{24}g_V^2\mu_1^2 \frac{1}{3\sqrt5}\left[2(p'^2+p^2)t-p'p(3+t^2)\right] G(\rho,\omega,q_{0C},p',p,t)\Bigg\}.\label{eq:V1113}
		\end{align}
\end{widetext}

\begin{widetext}
	\allowdisplaybreaks
		The partial wave projected potentials for $D^*\bar{D}_2^{*}\to D^*\bar{D}_2^{*}$ (channels 14--15) are:
		\begin{align}
			V_{14,14}(p',p) &= 2\pi\int_{-1}^{1}dt\Bigg\{-\frac{gg''}{2f_\pi^2}\frac{3}{20} \left[4(p'^2+p^2)t-p'p(1+7t^2)\right]G(\pi,\eta,q_{0D},p',p,t)\nonumber\\
			&\quad + \frac{(h_1+h_2)^2}{2\Lambda_\chi^2 f_\pi^2} \frac{3}{20}p'p(p'^2+p^2-2p'pt)(1-t^2)G(\pi,\eta,q_{0C},p',p,t)\nonumber\\
			&\quad -g_\sigma g''_\sigma t G(m_\sigma,q_{0D},p',p,t)+ \frac{h_\sigma^{'2}}{f_\pi^2}\frac{1}{60}[-12(p'^2+p^2)t+p'p(23+t^2)]G(m_\sigma,q_{0C},p',p,t)\nonumber\\
			&\quad +g_V^2\left[\frac{1}{4}\beta\beta_2 t+\lambda\lambda_2\frac{3}{20}[6(p'^2+p^2)t+p'p(1-13t^2)]\right]G(\rho,\omega,q_{0D},p',p,t)\nonumber\\
			&\quad +\frac{g_V^2}{4}\Bigg[\frac{1}{6}(\zeta_1-\mu_1 q_{0C})^2 t-\mu_1^2\frac{1}{60}[-12(p'^2+p^2)t+p'p(23+t^2)]\Bigg] G(\rho,\omega,q_{0C},p',p,t)\Bigg\},\label{eq:V1414}\\[5pt]
			V_{15,15}(p',p) &= 2\pi\int_{-1}^{1}dt\Bigg\{\frac{gg''}{2f_\pi^2}\frac{1}{20} \left[-8(p'^2+p^2)t+p'p(7+9t^2)\right]G(\pi,\eta,q_{0D},p',p,t)\nonumber\\
			&\quad - \frac{(h_1+h_2)^2}{2\Lambda_\chi^2 f_\pi^2} \frac{1}{20}\Big[4(p'^4+p^4)t+2p'^2 p^2 t(5+7t^2) -(p'^3 p+p' p^3)(1+15t^2)\Big]G(\pi,\eta,q_{0C},p',p,t)\nonumber\\
			&\quad-g_\sigma g''_\sigma t G(m_\sigma,q_{0D},p',p,t)- \frac{h_\sigma^{'2}}{f_\pi^2}\frac{1}{20}[8(p'^2+p^2)t-p'p(7+9t^2)]G(m_\sigma,q_{0C},p',p,t)\nonumber\\
			&\quad +g_V^2\left[\frac{1}{4}\beta\beta_2 t+\lambda\lambda_2\frac{1}{20}[2(p'^2+p^2)t+p'p(7-11t^2)]\right]G(\rho,\omega,q_{0D},p',p,t)\nonumber\\
			&\quad -\frac{g_V^2}{4}\Bigg[\frac{1}{2}(\zeta_1-\mu_1 q_{0C})^2 t-\mu_1^2\frac{1}{20}[8(p'^2+p^2)t-p'p(7+9t^2)]\Bigg]G(\rho,\omega,q_{0C},p',p,t)\Bigg\},\label{eq:V1515}\\[5pt]
			V_{14,15}(p',p) &= 2\pi\int_{-1}^{1}dt\Bigg\{\frac{gg''}{2f_\pi^2}\frac{\sqrt{3}}{10} \left[-2(p'^2+p^2)t+p'p(3+t^2)\right]G(\pi,\eta,q_{0D},p',p,t)\nonumber\\
			&\quad + \frac{h_\sigma^{'2}}{f_\pi^2}\frac{1}{5\sqrt{3}}[2(p'^2+p^2)t-p'p(3+t^2)] G(m_\sigma,q_{0C},p',p,t)\nonumber\\
			&\quad +g_V^2\left[\lambda\lambda_2\frac{\sqrt3}{10}[-2(p'^2+p^2)t+p'p(3+t^2)]\right]G(\rho,\omega,q_{0D},p',p,t)\nonumber\\
			&\quad -\frac{g_V^2}{4}\left[\mu_1^2\frac{1}{5\sqrt3}[2(p'^2+p^2)t-p'p(3+t^2)]\right] G(\rho,\omega,q_{0C},p',p,t)\Bigg\}. \label{eq:V1415}
		\end{align}
\end{widetext}

\subsection{Appendix D: Decay widths of $D_0^*$, $D_1^\prime$, $D_1$, and $D_2^*$}\label{appendix:width}

In this appendix, we assume that the total widths of the unstable mesons $D_0^*$, $D_1^\prime$, $D_1$, and $D_2^*$ are dominated by their two-body strong decays into the $D^{(*)}\pi$ channels. Given the large phase space available for these decays, isospin symmetry breaking effects are expected to be negligible. Consequently, based on isospin symmetry, the partial widths satisfy the relation $\Gamma(B\to D^{(*)}\pi^+) = 2\Gamma(B\to D^{(*)}\pi^0)$, where $B$ denotes the initial unstable meson. The explicit formulas for the decay widths into the charged pion mode ($\pi^+$) are given by:
\vspace{-1.0em} 
\begin{align}
	\Gamma(D_0^*\to D\pi^+) &= \frac{f''}{2\pi f_\pi^2}\frac{m_D}{m_{D_0^*}}k_{\pi}E_\pi^2,\label{eq:width D0star}\\
	\Gamma(D'_1\to D^*\pi^+) &= \frac{f''}{2\pi f_\pi^2}\frac{m_{D^*}}{m_{D'_1}}k_{\pi}E_\pi^2,\\
	\Gamma(D_1\to D^*\pi^+) &= \frac{1}{6\pi f_\pi^2}\left(\frac{h_1+h_2}{\Lambda_\chi}\right)^2\frac{m_{D^*}}{m_{D_1}}k_\pi^5,\label{eq:width D1}\\
	\Gamma(D_2^*\to D\pi^+) &= \frac{1}{15\pi f_\pi^2}\left(\frac{h_1+h_2}{\Lambda_\chi}\right)^2\frac{m_{D}}{m_{D_2^*}}k_\pi^5, \label{D2star width1}\\
	\Gamma(D_2^*\to D^*\pi^+) &= \frac{1}{10\pi f_\pi^2}\left(\frac{h_1+h_2}{\Lambda_\chi}\right)^2\frac{m_{D^*}}{m_{D_2^*}}k_\pi^5. \label{D2star width2}
\end{align}
where $k_\pi$ is the momentum of the pion in the rest frame of the decaying particle $B$, defined as:
\begin{align}
	k_{\pi} &= \frac{1}{2E_{D_0^*}}\lambda^{1/2}(m_{B}^2,m_{D^{(*)}}^2,m_\pi^2),\nonumber\\
	\lambda(a,b,c) &= a^2+b^2+c^2-2ab-2bc-2ac.
\end{align}
Here, $M_B$ represents the effective mass (or running energy $\sqrt{s}$) of the unstable particle $B$ (e.g., $E_{D_0^*}$ as defined in Eq.~\eqref{eq:width_sup}) when considering self-energy corrections.

	\bibliography{tot.bib}

\end{document}